\documentclass[preprint]{aastex}

\usepackage{graphicx}
\usepackage{array}
\usepackage{natbib}
\usepackage{amssymb}
\usepackage{lscape}

\begin{document}

\title{\textit{Spitzer} Photometry of $\sim1$\,Million Stars in M\,31 and 15 Other Galaxies \altaffilmark{1}}

\author{Rubab~Khan\altaffilmark{2,3}}

\altaffiltext{1}{Based on observations made with the {\it Spitzer} Space Telescope, 
which is operated by the Jet Propulsion Laboratory, California Institute of Technology 
under a contract with NASA.}

\altaffiltext{2}{JWST Fellow, NASA Postdoctoral Program, NASA Goddard Space Flight Center, MC 665,
8800 Greenbelt Road, Greenbelt, MD 20771}

\altaffiltext{3}{Department of Astronomy, Box 351580, University of Washington, Seattle, WA 98195; rubab@uw.edu}

\shorttitle{\textit{Spitzer} Catalog of $16$ Galaxies}

\shortauthors{Khan et al. 2015(b)}

\begin{abstract}
\label{sec:abstract}
We present \textit{Spitzer} IRAC $3.6-8\,\micron$ and MIPS $24\,\micron$ point-source 
catalogs for M\,$31$ and $15$ other mostly large, star forming galaxies at distances 
$\sim3.5-14$\,Mpc including M\,$51$, M\,$83$, M\,$101$ and
NGC\,$6946$. These catalogs contain $\sim1$\,million sources including $\sim859,000$ in M\,31 
and $\sim116,000$ in the other galaxies.
They were created following the procedures described in \citet{ref:Khan_2015b} through a 
combination of point spread function (PSF) fitting and aperture photometry. These data products constitute a 
resource to improve our understanding of the 
IR-bright ($3.6-24\,\mu$m) point-source populations in crowded extragalactic stellar fields and 
to plan observations with the James Webb Space Telescope.
\end{abstract} 

\maketitle

\section{Introduction}
\label{sec:intro}

The Infrared Array Camera~\citep[IRAC,][]{ref:Fazio_2004} and the Multiband 
Imaging Photometer~\citep[MIPS,][]{ref:Rieke_2004} instruments aboard the 
{\it Spitzer} Space Telescope \citep[{\it Spitzer},][]{ref:Werner_2004}
have collected a vast archive of mid-infrared (mid-IR) imaging data. 
This resource makes it feasible to  
identify and characterize mid-IR luminous stars in the crowded 
and dusty disks of large star forming galaxies despite difficulties 
due to IR emission from interstellar dust, blending and background contamination.
In \citet{ref:Khan_2010} we published the first ever mid-IR
point-source catalogs for galaxies significantly beyond the Local Group ($\gtrsim1.9$\,Mpc). 
In \citet{ref:Khan_2015b} we used  archival IRAC and MIPS images of seven galaxies 
at $\sim1-4$\,Mpc to catalog $\sim300,000$ stars, that were used to identify an emerging 
class of high mass ($>25 M_\odot$) post-main-sequence stars \citep{ref:Khan_2015a}.
Here we present photometric inventories of the mid-IR point sources in 
the IRAC $3.6\,\micron$, $4.5\,\micron$, $5.8\,\micron$ and $8\,\micron$ as well as MIPS $24\,\micron$ 
images of the Andromeda galaxy (M\,$31$) and $15$ other nearby ($\lesssim14\,$Mpc) galaxies 
beyond the Local Group. Our key motivation here is to facilitate targeted follow-up 
of individual objects and observation planning of IR-bright extragalactic stellar populations 
in the upcoming era of the James Webb Space Telescope \citep[JWST,][]{ref:Gardner_2006} 
and the Wide-Field InfrarRed Survey Telescope \citep[WFIRST,][]{ref:Spergel_2015}.

As the nearest major spiral galaxy to the Milky Way, the Andromeda galaxy (M\,$31$) have been 
extensively observed over the years from both ground and space based observatories 
\citep[e.g.,][]{ref:Baade_1944,ref:deVaucouleurs_1958,ref:Massey_2006,ref:Johnson_2012}.
The Panchromatic Hubble Andromeda Treasury \citep{ref:Dalcanton_2012} mapped roughly a third of M\,$31$'s star forming disk, 
using 6 filters covering from the ultraviolet through the near-infrared (near-IR) to produce the most detailed 
picture of resolved extragalactic stellar populations in a galaxy. However, public availability of mid-IR stellar 
catalogs of this galaxy is very limited. \citet{ref:Mould_2008} performed mid-IR photometry 
of point sources on {\it Spitzer} IRAC and MIPS images of M\,$31$. However, while 
their paper shows mid-IR color magnitude diagrams containing seemingly many hundred thousand sources,
they published only a small fraction ($\sim500-900$ sources at various bands) of the catalog, consisting 
of the brightest sources in the field.
In this paper, we present an extensive mid-IR point-source catalog of M\,$31$ consisting of 
$\sim859,000$ sources, covering the entirety of M\,$31$'s disk including the accompanying M\,$32$ and M\,$110$ galaxies. 

When selecting the other $15$ galaxies, we concentrated on those with 
higher recent star formation rate (SFR), as these would have large numbers 
of short lived, massive, evolved 
mid-IR bright stars, and we cataloged $\sim116,000$ stars in these 
galaxies. These catalogs include the highly star-forming galaxies M\,$83$, NGC\,$6946$, M\,$101$ and M\,$51$ 
(M\,$51$a and M\,$51$b) which enabled the first-ever identification of
extragalactic candidate analogs of the Galactic stellar behemoth $\eta$\,Carinae \citep{ref:Khan_2015c}.
We selected these $15$ galaxies to span a range of distances and SFRs ($\sim3.5-14$\,Mpc and 
$\lesssim0.1-\sim1 M_\odot /$year, see Table\,\ref{tab:stats}), and currently there are no public mid-IR stellar catalogs 
for $13$ of these galaxies. \citet{ref:Williams_2015} published a mid-IR bright source catalog of M\,$83$ including 
Spitzer $3.6$ and $4.5\,\micron$ band measurements for $<4,000$ objects, while the 
M\,83 catalog presented here contains Spitzer $3.6$, $4.5$, $5.8$, $8$ and $24\,\micron$ band measurements 
for $\sim23,000$ sources. Likewise, \citet{ref:Khan_2010} reported two Spitzer band measurements of $<6,000$ objects 
in NGC\,$6946$ whereas the catalog for this galaxy presented here contains five {\it Spitzer} band measurements for $\sim16,000$ sources.

For M\,$31$, we used the IRAC $3.6\,\micron$, $4.5\,\micron$, $5.8\,\micron$ and $8\,\micron$ mosaics 
produced by \citet{ref:Mould_2008} and the MIPS $24\,\micron$ mosaic produced by \citet{ref:Gordon_2006}. 
For the other galaxies, we used the IRAC and MIPS mosaics produced by the \textit{Spitzer} Infrared Nearby 
Galaxies Survey~\citep[SINGS,][]{ref:Kennicutt_2003} and the Local Volume Legacy Survey~\citep[LVL,][]{ref:Dale_2009}. 
We utilize the full mosaics available for each galaxy. The M\,$31$ mosaics (covering 
$\sim4.2$ square degrees) are constructed 
from many individual exposures whereas each of the more distant galaxy images 
from the LVL and SINGS archives are usually combinations of two slightly 
offset images. We do not take advantage of the uncertainty maps and 
assume mean noise properties, since in our experience the dominant source 
of flux uncertainty in this context are related to crowding, which varies
significantly across face of the target galaxies.
In what follows, we summarize our methodology 
(Section\,\ref{sec:photo}, see \citet{ref:Khan_2015b} for details), and 
discuss properties of the catalogs and color-magnitude distributions 
(Section\,\ref{sec:cats}). 

\section{Photometry}
\label{sec:photo}

We obtained the photometric 
measurements at various wavelengths and combined them to construct the 
point-source catalogs following the procedures established in \citet{ref:Khan_2015b}. 
We implement a strict detection criteria by requiring $>3\sigma$ 
detection of all cataloged sources at $3.6\,\micron$ and $4.5\,\micron$. 
We then complement those measurements 
at the $5.8\,\micron$, $8.0\,\micron$ and $24\,\micron$ bands through a combination of 
point spread function (PSF) fitting photometry and 
aperture photometry. For all objects that do not have a $>3\,\sigma$ detection 
at these three longer wavelengths, we estimate their $3\sigma$ flux upper
limits in those bands.

First we select all sources detected through PSF fitting photometry at $>3\sigma$
in both the $3.6\,\micron$ and $4.5\,\micron$ images within a $1$\,pixel
matching radius as point sources. Next, 
we search for $>3\sigma$ detections of these point sources 
in the $5.8\,\micron$ and $8.0\,\micron$ images within the same matching 
radius. If no counterpart is found, we attempt 
to measure the flux at the location of the $3.6/4.5 \micron$ point source through PSF fitting, 
and failing that, through aperture photometry. For the MIPS 
$24\,\micron$ images, we only use aperture photometry 
due to the much lower resolution and larger PSF size compared to the IRAC 
images\footnote[4]{Mean full width half-max (FWHM) of the cryogenic IRAC PSFs 
are $1\farcs66$, $1\farcs72$, $1\farcs88$ and $1\farcs98$, and the
MIPS $24\,\micron$ PSF FWHM is 
$5\farcs9$.}$^{,}$\footnote[5]{\tt http://irsa.ipac.caltech.edu/data/SPITZER/docs/irac/iracinstrumenthandbook/5/}$^{,}$\footnote[6]{\tt http://irsa.ipac.caltech.edu/data/SPITZER/docs/mips/mipsinstrumenthandbook/50/}. 
Finally, for all objects that do not have a $>3\,\sigma$ detection 
at $5.8\,\micron$, $8.0\,\micron$ and $24\,\micron$, we estimate the $3\sigma$ flux upper
limits. The fluxes and upper limits are transformed to Vega-calibrated magnitudes using 
the flux zero-points
and aperture corrections provided in the \textit{Spitzer} Data Analysis 
Cookbook\footnote[7]{\tt http://irsa.ipac.caltech.edu/data/SPITZER/docs/dataanalysistools/}.

We used the DAOPHOT/ALLSTAR PSF-fitting and photometry package~\citep{ref:Stetson_1992} 
to construct the PSFs, to identify the $>3\sigma$ sources and to measure their 
flux at all $4$ IRAC bands. We used the IRAF\footnote[8]{IRAF is distributed by the 
National Optical Astronomy Observatory, which is operated by the Association of 
Universities for Research in Astronomy (AURA) under cooperative agreement with 
the National Science Foundation.} ApPhot/Phot tool for performing aperture photometry
for all IRAC bands and the MIPS $24\,\micron$ band.
For the four IRAC bands, we use an extraction aperture of $2\farcs4$, a local background 
annulus of $2\farcs4 - 7\farcs2$ and aperture corrections of $1.213$, $1.234$, $1.379$, 
and $1.584$ respectively.
For the MIPS $24\,\micron$ band, we use an extraction aperture of $3\farcs5$, a local background 
annulus of $6\farcs - 8\farcs$ and an aperture correction of $2.78$.
We estimate the local background using a $2\sigma$ outlier rejection procedure in 
order to exclude sources located in the local sky annulus and correct for the 
excluded pixels assuming a Gaussian background distribution. We determine 
the $3\sigma$ flux upper limit for each aperture location using the 
local background estimate.

We present the results of our mid-IR photometric survey following 
the same format as the catalogs published in \citet{ref:Khan_2015b}. 
Tables\,$2-17$ list the coordinates (J$2000.0$; RA and Dec) of the 
point sources followed by their Vega calibrated apparent magnitudes ($m_\lambda$), the associated 
$1\,\sigma$ uncertainties ($\sigma_\lambda$) and (for the $3.6-8.0\,\micron$ bands) the differences 
between the PSF and aperture photometry magnitudes ($\delta_\lambda$).
For the $5.8\,\micron$, $8.0\,\micron$ and $24\,\micron$ bands, 
$\sigma_\lambda=99.99$ implies that the associated photometric measurement is a
$3\,\sigma$ flux upper limit, and $m_\lambda=99.99$ (as well as 
$\sigma_\lambda=99.99$) indicates that no reliable photometric measurement could be 
obtained for that location. For the IRAC bands, $\delta_\lambda=99.99$ implies 
that one or both of the associated photometric
measurements did not yield a $>3\sigma$ flux measurement.

Large mismatches between the two (PSF-fitting and aperture) measurements, specially when\\ 
$|\delta_\lambda| >> \sigma_\lambda $, are a 
good indicator of when crowding is significantly effecting the photometry 
and can be useful as an alternative estimate of photometric uncertainty. 
While PSF-fitting photometry may 
be generally preferable for crowded field photometry
where possible, the $\delta_\lambda$ values would let one revert to using the aperture 
photometry measurements instead.
Large $\delta_\lambda$ values associated with seemingly bright sources are also 
indicative of contamination due to saturated foreground objects being resolved into 
multiple bright sources by the PSF-fitting point source detection procedure 
\citep[foreground giants would have m$\lesssim7$, e.g.,][]{ref:McQuinn_2007}. 
This is a major source of false positives, specially for M\,$31$, as its large field 
of view contains numerous foreground objects. Indeed, our attempt to identify evolved 
dust-obscured very high-mass stars 
($M_{ZAMS}\gtrsim25 M_\odot$) in M\,$31$ following the selection criteria described 
in \citet{ref:Khan_2013} picked up many such spurious sources due to their apparently
peculiar spectral energy distributions (SEDs).

\section{Discussion}
\label{sec:cats}

\begin{figure}
\begin{center}
\includegraphics[angle=0,width=150mm]{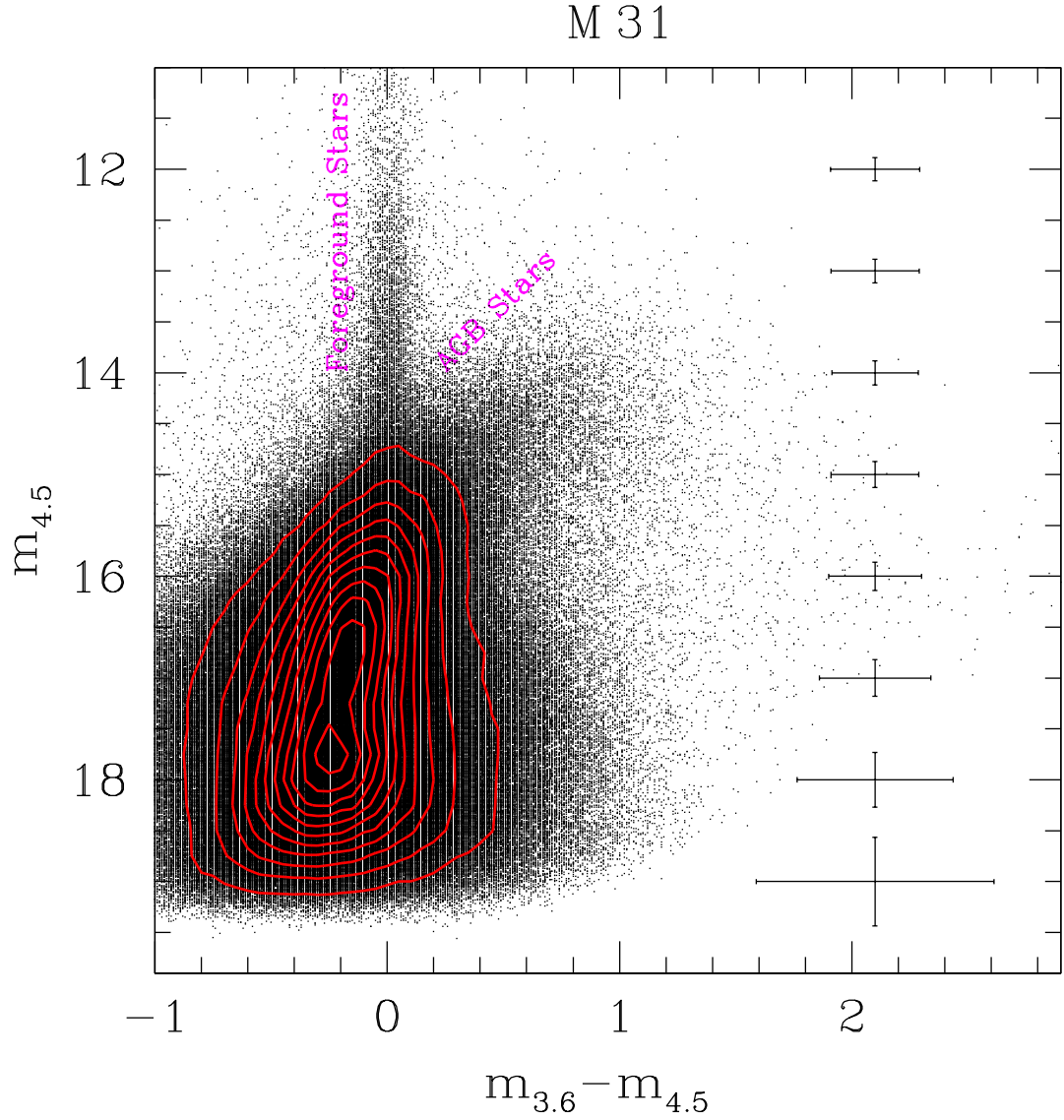}
\end{center}
\caption{The $m_{4.5}$\,vs.\,$m_{3.6}-m_{4.5}$ color magnitude diagram (CMD) for the 
cataloged sources in M\,31. The red lines 
represent isodensity contours, and the error bars show mean $1\sigma$ color and 
magnitude uncertainties for cataloged sources in $1$\,magnitude bins.}
\label{fig:cmd1}
\end{figure}

\begin{figure}
\begin{center}
\includegraphics[angle=0,width=150mm]{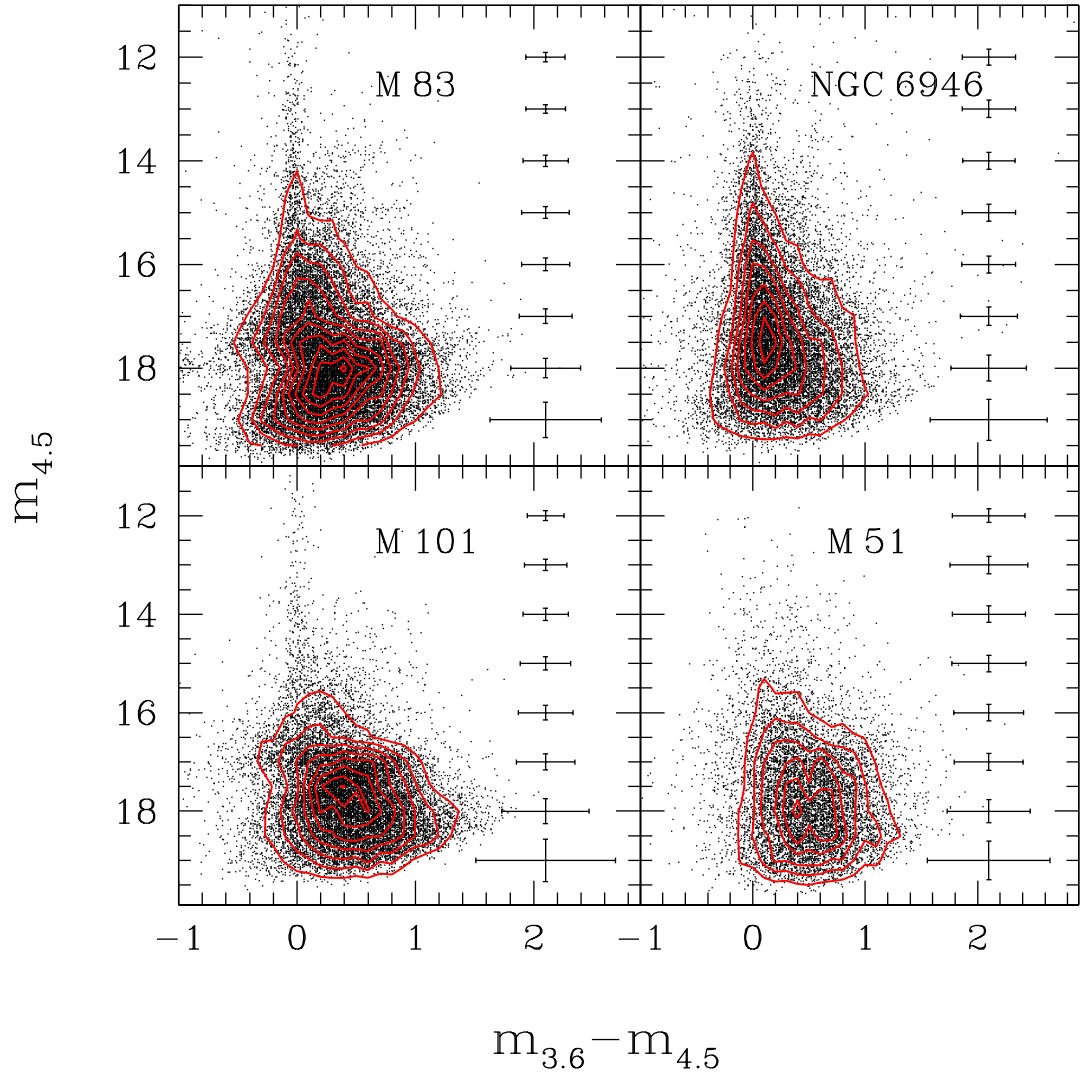}
\end{center}
\caption{Same as Figure\,\ref{fig:cmd1} for the galaxies M\,51, M\,83, 
M\,101 and NGC\,6946.}
\label{fig:cmd4}
\end{figure}

Figure\,\ref{fig:cmd1} shows the $m_{4.5}$\,vs.\,$m_{3.6}-m_{4.5}$ 
color magnitude diagram (CMD) for M\,$31$ and Figure\,\ref{fig:cmd4} 
shows the same for the galaxies M\,$83$, NGC\,$6946$, M\,$101$ and M\,$51$ 
which have the highest SFR among all the galaxies surveyed. The
$1\sigma$ color and magnitude uncertainties  
indicate that the horizontal extent of the 
CMDs are largely a result of color uncertainties.
The blue-ward extent of 
the M\,$31$ CMD is consistent with, e.g, the comparable CMDs of M\,$33$ shown on Fig.\,$14$ 
of \citet{ref:McQuinn_2007} and Fig.\,$4$ of \citet{ref:Khan_2015b},
and it contains a larger fraction of blue sources than   
the $15$ galaxies beyond the local group cataloged here. As these galaxies are between factors of 
$\sim4.5$ (NGC\,$3077$ at $3.7$\,Mpc) and $\sim18$ (NGC\,$3184$ at $14.4$\,Mpc) farther away 
than M\,31 (at $0.78$\,Mpc), in M\,$31$ we identify intrinsically fainter and lower 
mass stars with relatively bluer colors. These include O- and C-rich Asymptotic Giant Branch (AGB) stars 
\citep[e.g.,][]{ref:Bolatto_2007} and possibly some Red Giant Branch (RGB) stars
($m_{4.5}\lesssim18$ in M\,$31$, e.g., \citealp{ref:Blum_2006,ref:Boyer_2015})
as well as the  more evolved and more luminous (massive) stars 
with warm circumstellar dust which have redder mid-IR colors (M\,$31$ is known
to have some young massive stars, e.g., \citealp{ref:Lewis_2015,ref:Massey_2016}).

All normal stars have the same mid-IR color in the first two IRAC bands,
because of the Rayleigh-Jeans tails of their spectra,
and we see this as a sequence of foreground dwarfs with $m_{3.6}-m_{4.5}\simeq0$ 
on the M\,$31$ CMD, as well as a ``plume'' of bright and red 
extreme Asymptotic Giant Branch (ex-AGB) stars 
\citep{ref:Thompson_2009,ref:Khan_2010,ref:Boyer_2015} at $m_{4.5}\simeq13-16$. 
This feature is not as prominent on the M\,$31$ CMD when compared 
to the tight stream of ex-AGB stars visible on the CMD of M\,$33$, 
which has significantly higher ($\gtrsim10\times$) specific star formation rate 
\citep[e.g., see][for a detailed discussion of M\,$31$'s recent star formation history]{ref:Lewis_2015}
and thus a larger number of younger massive stars per unit stellar mass, although it is 
still a prominent feature when compared to CMDs of even lower mass/SFR galaxies such as NGC\,$6822$
(see Fig.\,$4$ of \citealp{ref:Khan_2015b} for M\,$33$ and NGC\,$6822$ mid-IR CMDs). 

However, quasars also have this color \citep[e.g.,][]{ref:Stern_2005}, 
as do star forming galaxies with strong PAH emission at $8.0\micron$ 
(e.g., the SED models in \citealp{ref:Assef_2010}),
and the ex-AGB stars are far less 
noticeable amid background contaminants 
on the more distant galaxy CMDs. 
Although these galaxies have smaller effective survey areas and 
do not have more background contamination per unit area than M\,$31$, their 
greater distance modulus ($\mu$) means that stars in those galaxies 
have larger apparent magnitudes. As a result, the evolved stellar 
populations in those galaxies are effectively buried among background 
sources on the CMDs. For example, the tip of the AGB branch that is at 
$m_{4.5}\simeq13$ on the M\,$31$ CMD ($\mu\simeq24.5$, Figure\,\ref{fig:cmd4}) would be at  
at $m_{4.5}\simeq17$ on the M\,$83$ CMD ($\mu\simeq28.3$, Figure\,\ref{fig:cmd4}). 
Given the rarity of ex-AGB stars 
\citep[e.g.,][]{ref:Thompson_2009,ref:Khan_2010,ref:Boyer_2015} it 
is very likely that most of the very red ($m_{3.6}-m_{4.5}\gtrsim1$) sources 
we identify are background contaminants. 
However, verifying the nature of individual sources would 
require mid-IR variability study and/or construction of extended 
multi-wavelength SEDs on a case-by-case basis \citep[e.g.,][]{ref:Khan_2013}. 
Indeed, only a rare few cataloged mid-IR sources in these distant galaxies 
would be relevant in the context of studying properties of individual stars
--- but that is what makes them very interesting to analyze
\citep[e.g.,][]{ref:Khan_2015c}.

\begin{figure}
\begin{center}
\includegraphics[angle=0,width=150mm]{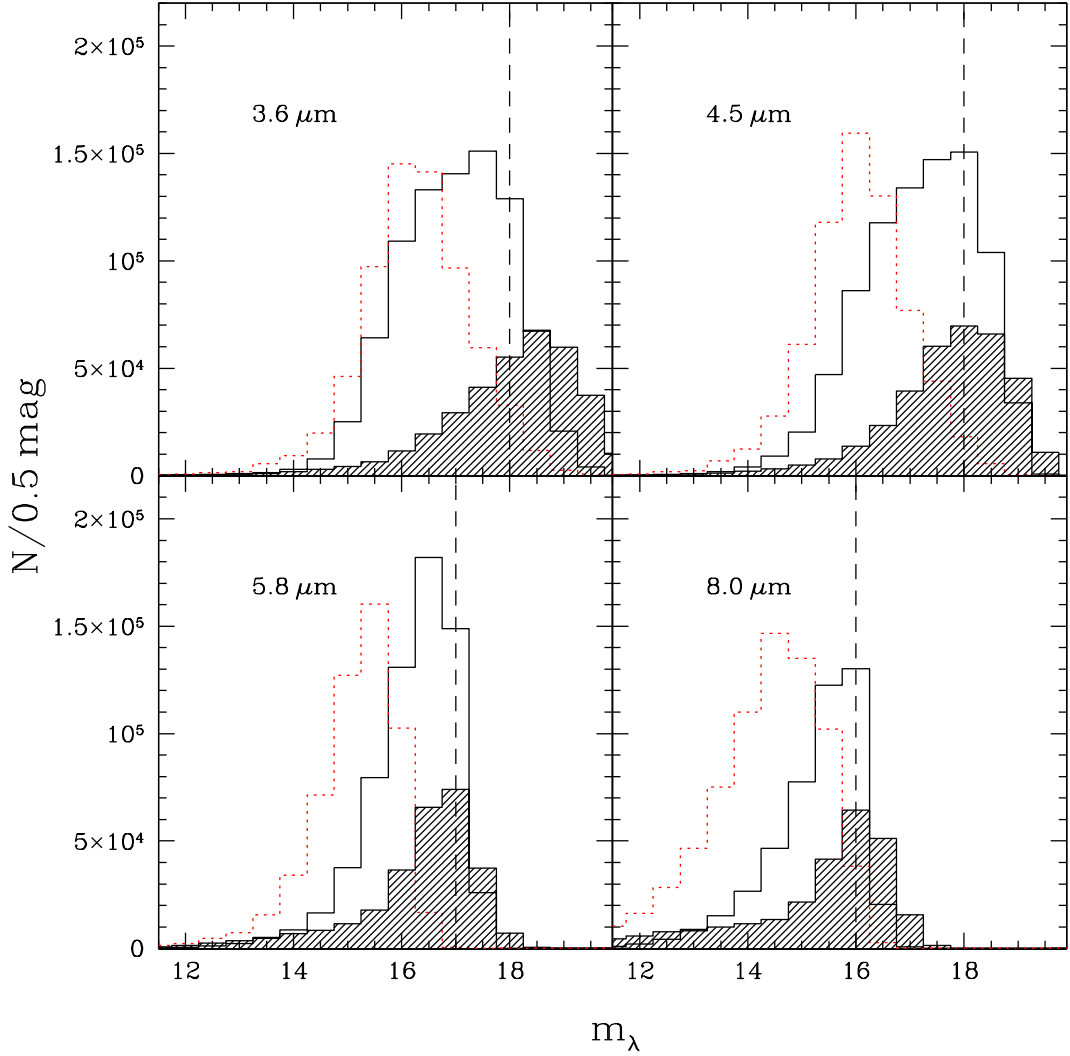}
\end{center}
\caption{Apparent magnitude histograms for all cataloged sources in M\,$31$ 
and in the other $15$ galaxies (shaded regions), with the latter scaled up 
for clarity by a factor of $3$. The dotted lines show the apparent-magnitude 
histograms of the SDWFS catalog sources, scaled up  for clarity by a factor of $30$
for $m_{3.6}$ and $m_{4.5}$, and by a factor of $50$ for $m_{5.8}$ and $m_{8.0}$.}
\label{fig:mHist}
\end{figure}

Figure \ref{fig:mHist} 
shows the apparent magnitude histograms of all sources in the catalog of M\,$31$ 
(clear region) as well as in the other galaxies (shaded region). 
For a qualitative comparison, we also show the magnitude 
histogram for all sources in a $6$\,deg$^2$ region of the NOAO Bootes Field produced 
from the {\it Spitzer} Deep Wide Field Survey \citep[SDWFS,][]{ref:Ashby_2009} data (dotted line).
The SDWFS catalog can be largely considered ``empty'' as in 
most sources being background galaxies and quasars, with only a small fraction 
being foreground stars \citep[e.g., see][]{ref:Kozlowski_2016}.
Figures\,\ref{fig:mHist} shows that our catalogs are $\gtrsim1$\,mag deeper than the SDWFS catalog. 
It is worth noting that our catalogs
simply inventory all the sources present on the image mosaics
and we do not attempt to distinguish between sources physically associated with the 
galaxies and unrelated foreground and background contaminants. 
We cannot claim completeness at any magnitude limit, rather can 
only infer that the observed luminosity 
function turns over at a certain magnitude while the intrinsic one continues 
rising as the catalogs become increasingly incomplete for fainter sources.
Overall, Figure\,\ref{fig:mHist} qualitatively implies that our source lists 
become significantly incomplete at $m_{3.6}\gtrsim18$, $m_{4.5}\gtrsim18$, 
$m_{5.8}\gtrsim17$ and $m_{8.0}\gtrsim16$.

As we emphasized in \citet{ref:Khan_2015b}, 
point-source catalogs of the 
inherently crowded galaxy fields that we are surveying are 
bound to be crowding (confusion) limited, not just magnitude limited. 
While Figure\,\ref{fig:mHist} empirically demonstrates that our source detection 
peaks at a certain magnitude and then falls off rapidly, it is likely that 
incompleteness is affecting even the bright-star counts in crowded regions, increasing towards and 
through the peak. The depth and completeness of the catalogs vary across each galaxy 
between, e.g., the centers 
of galaxies compared to their outer regions or in dusty star clusters compared to 
more sparsely populated regions
as a function of crowding, and they can only be characterized locally for small regions. 
Performing a conventional efficiency determination test 
through addition of randomly distributed artificial objects in the images therefore 
would lead us to either overestimate or
underestimate the efficiency. For such a study to be truly useful, it would require a 
proper ``star-star correlation function'' to be employed for spatial distribution of 
artificial stars. 
Also, while we execute the point source detection procedures in 
individual bands, a source is included in the catalog only if it is 
independently identified as a point-source in both the $3.6$ and $4.5\micron$ bands 
at least at a $3\sigma$ level by PSF fitting. Any meaningful statistical test in this 
context therefore would also 
need to account for stellar SED variations in the mid-IR 
to test multi-band catalog completeness for a particular region of interest.

\begin{figure}
\begin{center}
\includegraphics[angle=0,width=150mm]{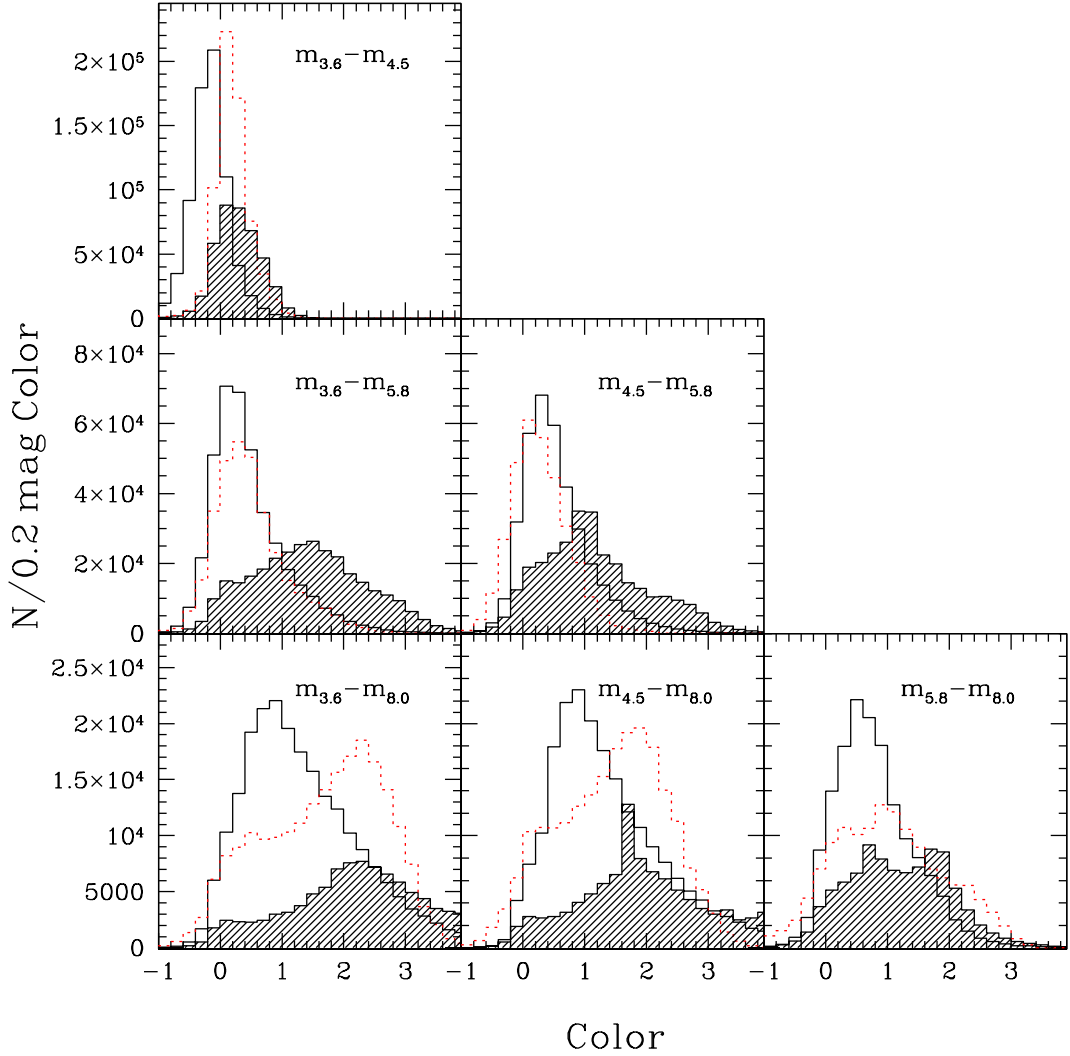}
\end{center}
\caption{Mid-IR color histograms for all cataloged sources in M\,31 
and in the other $15$ galaxies (shaded regions)
with $1\,\sigma$ uncertainty in color $\lesssim0.2$, with the latter scaled up 
for clarity by factors of $5$ (first and second rows) and $2$ (bottom row).
The dotted lines show the mid-IR color
histograms of the SDWFS catalog sources, scaled-up  for clarity by factors of
$30$ (first and second rows) and $15$ (bottom row).}
\label{fig:cHist}
\end{figure}

Figure\,\ref{fig:cHist} shows the mid-IR color histograms of all sources in the catalogs
with $1\,\sigma$ uncertainty in color $\lesssim0.2$, 
following the same representation as Figure \ref{fig:mHist}. 
As discussed earlier in this section, the $m_{3.6}-m_{4.5}$ color distribution 
of M\,$31$ is skewed blue-ward compared to the other galaxies.
The $m_{3.6}-m_{5.8}$ and $m_{4.5}-m_{5.8}$ color distributions (middle row of 
Figure\,\ref{fig:cHist}) of these more distant galaxies peak at 
$>1$\,mag redder relative to M\,$31$ and the Bootes field, indicating
that their $5.8\,\micron$ flux may be dominated by PAH emissions, which is a
common feature of massive star-forming regions and star clusters 
\citep[e.g.,][]{ref:Churchwell_2006} created by 
strong stochastic emission from PAH molecules \citep[e.g.,][]{ref:Whelan_2011} excited by 
UV radiation from O- and B-type stars (see \citealp{ref:Wood_2008} for a detailed 
treatment of this topic). Their 
$m_{3.6}-m_{8.0}$, $m_{4.5}-m_{8.0}$ and $m_{5.8}-m_{8.0}$ color distributions 
generally match those of the Bootes field but are redder than M\,$31$ 
(bottom row of Figure\,\ref{fig:cHist}), consistent with 
significant extragalactic contamination (see Fig.\,$5$ of \citealp{ref:Khan_2015b} 
for mid-IR CMDs of the SDWFS sources). 

It is important to highlight here that the color histograms do not 
include sources for which we could only measure a flux upper-limits at 
the $5.8$ and/or $8.0\micron$ bands. Since the catalogs 
list sources that have $>3\sigma$ detections at the $3.6$ 
and $4.5\micron$, the middle and bottom rows of Figure\,\ref{fig:cHist}
are inherently biased toward redder sources, i.e., those with 
$>3\sigma$ detections at the two longer wavelength 
bands as well as the two shorter ones. This can exclude relatively bluer 
sources such as foreground dwarfs as well as O- and C-rich AGB stars in the 
targeted galaxies that are 
intrinsically less luminous at the longer wavelengths.
A more rigorous pursuit 
of this topic requires studying near-IR to mid-IR color separations 
of the cataloged sources, e.g., as done for the LMC 
by \citet{ref:Blum_2006} utilizing 2MASS data. However, 2MASS is not deep enough to study 
stellar populations in other galaxies (even M\,31's distance modulus is $\sim6$\,mags 
larger than the LMC's) and one would need (e.g.) WFIRST's Wide Field Instrument 
\citep[WFI,][]{ref:Spergel_2015} near-IR imaging 
data for this purpose.

These catalogs are a resource to improve our understanding of the 
$3.6-24\,\micron$ bright point-source populations in crowded extragalactic fields and they
are also an archive for studying future mid-IR transients. 
The JWST's 
Near-IR Spectrograph~\citep[NIRSpec,][]{ref:Dorner_2016}
and Mid-IR Instrument~\citep[MIRI,][]{ref:Rieke_2015}
will cover the $\sim1-5\,\micron$ and
$\sim5-28\,\micron$ wavelength 
ranges respectively, but the JWST's small field of view and anticipated over-subscription practically means that 
these catalogs will continue to be the most detailed listing of mid-IR source properties in 
nearby galaxies in the near future. These $3.6-24\,\micron$ point-source catalogs 
can be very useful to identify scientifically interesting sources for photometric and spectroscopic
follow-up with NIRSpec and MIRI in general. They create a pathway for the 
exploration of extragalactic evolved stellar populations as well as other mid-IR bright 
sources with the JWST and WFIRST, 
making optimal and efficient use of these flagship observatories.

\acknowledgments
We thank the referee for helpful suggestions, 
Krzysztof Stanek, Christopher Kochanek and George Sonneborn for productive 
discussions, and Martha Boyer and Karl Gordon for providing the M\,31 image mosaics.
This work is based on observations made with the {\it Spitzer} Space Telescope, 
which is operated by the Jet Propulsion Laboratory, California Institute of Technology 
under a contract with the National Aeronautics and Space Administration (NASA). 
We extend our gratitude to the SINGS Legacy Survey 
and the LVL Survey for making their data publicly available. 
This research has made use of NED, which is 
operated by the JPL and Caltech, under contract with NASA and the HEASARC 
Online Service, provided by NASA's GSFC. 
RK is supported through a JWST Fellowship hosted by the Goddard Space Flight Center and 
awarded as part of the NASA Postdoctoral Program operated by the Oak Ridge Associated 
Universities on behalf of NASA.

\clearpage

\begin{appendix}
\label{appendix}

\begin{table}
\begin{center}
\begin{footnotesize} 
\caption{Catalog Statistics}
\label{tab:stats}
\begin{tabular}{lrrllclr}
\\
\hline 
\hline
\\
\multicolumn{1}{c}{Galaxy} &
\multicolumn{2}{c}{Galactic coor.} &
\multicolumn{2}{c}{Distance} &
\multicolumn{1}{c}{Data} &
\multicolumn{1}{c}{SFR\tablenotemark{a}} &
\multicolumn{1}{c}{Number}
\\
\multicolumn{1}{c}{} &
\multicolumn{1}{c}{{\it longitude}} &
\multicolumn{1}{c}{{\it latitude}} &
\multicolumn{1}{c}{(Mpc)} &
\multicolumn{1}{c}{{\it reference}} &
\multicolumn{1}{c}{Source} &
\multicolumn{1}{c}{$M_\odot /$year} &
\multicolumn{1}{c}{of Sources} 
\\
\\
\hline 
\hline
\\
M\,$31$                &  $121.174$  &  $-21.573$  & $0.78$  & \citet{ref:Stanek_1998}& {\it Note}\tablenotemark{b} & $0.7$\tablenotemark{c}    & $859,165$\\
NGC\,$3077$            &  $141.899$  &   $41.659$  & $3.7$   & \citet{ref:Jacobs_2009}     & LVL\tablenotemark{d}        & $0.076$    & $3,794$  \\
NGC\,$1313$            &  $283.359$  &  $-44.643$  & $4.4$   & \citet{ref:Jacobs_2009}     & LVL\tablenotemark{d}        & $0.316$    & $6,972$  \\
NGC\,$5236$            &  $314.584$  &   $31.973$  & $4.61$  & \citet{ref:Saha_2006}       & LVL\tablenotemark{d}        & $1.411$    & $23,331$ \\
NGC\,$4736$            &  $123.363$  &   $76.007$  & $4.66$  & \citet{ref:Jacobs_2009}     & SINGS\tablenotemark{e}      & $0.224$    & $10,264$ \\
NGC\,$4826$            &  $315.680$  &   $84.423$  & $4.66$  & \citet{ref:Jacobs_2009}     & SINGS\tablenotemark{e}      & $0.355$    & $5,137$  \\
NGC\,$5068$            &  $311.487$  &   $41.376$  & $5.45$  & \citet{ref:Herrmann_2008}    & LVL\tablenotemark{d}       & $0.524$    & $4,568$  \\
NGC\,$6946$            &   $95.719$  &   $11.673$  & $5.7 $  & \citet{ref:Sahu_2006}       & SINGS\tablenotemark{e}      & $2.289$    & $15,813$ \\
NGC\,$5474$            &   $64.301$  &   $22.933$  & $6   $  & \citet{ref:Rozanski_1994}   & SINGS\tablenotemark{e}      & $0.115$    & $991$    \\
NGC\,$5457$            &  $102.037$  &   $59.771$  & $6.43$  & \citet{ref:Shappee_2011}    & LVL\tablenotemark{d}        & $1.697$    & $16,291$ \\
NGC\,$45$              &   $55.903$  &  $-80.672$  & $6.6$   & \citet{ref:Jacobs_2009}     & LVL\tablenotemark{d}        & $0.245$    & $4,321$  \\
NGC\,$5194$            &  $104.851$  &   $68.561$  & $8   $  & \citet{ref:Ferrarese_2000}  & SINGS\tablenotemark{e}      & $1.512$    & $8,601$  \\
NGC\,$2903$            &  $208.711$  &   $44.540$  & $8.55$  & \citet{ref:Tully_2009}      & LVL\tablenotemark{d}        & $0.932$    & $5,579$  \\
NGC\,$925$             &  $144.885$  &  $-25.174$  & $9.3 $  & \citet{ref:Silbermann_1996} & SINGS\tablenotemark{e}      & $0.562$    & $4,217$  \\
NGC\,$3627$            &  $241.961$  &   $64.418$  & $10.5 $ & \citet{ref:Freedman_2001}   & LVL\tablenotemark{d}        & $1.022$    & $3,102$  \\
NGC\,$3184$            &  $178.336$  &   $55.638$  & $14.4 $ & \citet{ref:Ferrarese_2000}  & SINGS\tablenotemark{e}      & \dots    & $3,548$  
\\
\hline
\hline
\end{tabular}
\end{footnotesize} 
\tablenotetext{a}{H$\alpha$ luminosity from \citet{ref:Kennicutt_2008}are converted to star formation rate (SFR)
following Equation\,2 of \citet{ref:Kennicutt_1998}.}
\tablenotetext{b}{IRAC $3.6-8.0\,\micron$ from \citep{ref:Barmby_2006} and MIP $24\,\micron$ from \citet{ref:Gordon_2006}.}
\tablenotetext{c}{See \citet{ref:Lewis_2015} for a detailed discussion of M\,$31$'s recent star formation history.}
\tablenotetext{d}{Local Volume Legacy Survey~\citep[LVL,][]{ref:Dale_2009}.}
\tablenotetext{e}{\textit{Spitzer} Infrared Nearby Galaxies Survey \citep[SINGS,][]{ref:Kennicutt_2003}.}

\end{center}
\end{table}

\clearpage

\clearpage

\clearpage

\clearpage

\clearpage

\begin{landscape}

\begin{table}   
\begin{center}   
\begin{small}  
\caption{Catalog for $859,165$ Point Sources in M\,$31$} 
\label{tab:m31}   
\begin{tabular}{rrrrrrrrrrrrrrrrr}   
\\   
\hline  
\hline  
\multicolumn{1}{c}{RA} &
\multicolumn{1}{c}{Dec} &
\multicolumn{1}{c}{} &
\multicolumn{1}{c}{$m_{3.6}$} &
\multicolumn{1}{c}{$\sigma_{3.6}$} &
\multicolumn{1}{c}{$\delta_{3.6}$} &
\multicolumn{1}{c}{$m_{4.5}$} &
\multicolumn{1}{c}{$\sigma_{4.5}$} &
\multicolumn{1}{c}{$\delta_{4.5}$} &
\multicolumn{1}{c}{$m_{5.8}$} &
\multicolumn{1}{c}{$\sigma_{5.8}$} &
\multicolumn{1}{c}{$\delta_{5.8}$} &
\multicolumn{1}{c}{$m_{8.0}$} &
\multicolumn{1}{c}{$\sigma_{8.0}$} &
\multicolumn{1}{c}{$\delta_{8.0}$} &
\multicolumn{1}{c}{$m_{24}$} &
\multicolumn{1}{c}{$\sigma_{24}$} 
\\
\multicolumn{1}{c}{(deg)} &
\multicolumn{1}{c}{(deg)} &
\multicolumn{1}{c}{} &
\multicolumn{1}{c}{(mag)} &
\multicolumn{1}{c}{} &
\multicolumn{1}{c}{} &
\multicolumn{1}{c}{(mag)} &
\multicolumn{1}{c}{} &
\multicolumn{1}{c}{} &
\multicolumn{1}{c}{(mag)} &
\multicolumn{1}{c}{} &
\multicolumn{1}{c}{} &
\multicolumn{1}{c}{(mag)} &
\multicolumn{1}{c}{} &
\multicolumn{1}{c}{} &
\multicolumn{1}{c}{(mag)} &
\multicolumn{1}{c}{} 
\\ 
\hline  
\hline
\dots        & \dots       && \dots  & \dots  & \dots  & \dots  & \dots  & \dots  & \dots  & \dots  & \dots  & \dots  & \dots  & \dots  & \dots  & \dots  \\
 $ 11.48283$  &  $42.63817$   &&   $10.01$  &  $0.03$  &  $0.10$  &  $  9.68$  &  $0.04$  &  $-0.01$  &  $  9.57$  &  $0.03$  &  $-0.12$  &  $  9.54$  &  $0.03$  &  $0.03$  &  $  9.72$  &  $0.04   $  \\  
 $ 11.54072$  &  $41.57872$   &&   $10.46$  &  $0.13$  &  $0.83$  &  $  9.69$  &  $0.08$  &  $1.02$  &  $  6.79$  &  $0.01$  &  $99.99$  &  $  9.53$  &  $0.16$  &  $2.45$  &  $  6.18$  &  $0.01   $  \\  
 $ 10.56790$  &  $41.51002$   &&   $10.12$  &  $0.10$  &  $1.01$  &  $  9.70$  &  $0.06$  &  $0.86$  &  $  8.30$  &  $0.01$  &  $99.99$  &  $  8.32$  &  $0.03$  &  $0.09$  &  $  8.08$  &  $0.01   $  \\  
 $ 10.26228$  &  $41.57781$   &&   $10.00$  &  $0.04$  &  $0.06$  &  $  9.71$  &  $0.04$  &  $-0.01$  &  $  9.65$  &  $0.03$  &  $-0.05$  &  $  9.89$  &  $0.06$  &  $0.21$  &  $  9.71$  &  $0.04   $  \\  
 $ 10.01774$  &  $42.00484$   &&   $10.02$  &  $0.01$  &  $0.11$  &  $  9.71$  &  $0.02$  &  $0.01$  &  $  9.66$  &  $0.03$  &  $-0.18$  &  $  9.70$  &  $0.03$  &  $0.03$  &  $  9.61$  &  $0.02   $  \\  
\dots        & \dots       && \dots  & \dots  & \dots  & \dots  & \dots  & \dots  & \dots  & \dots  & \dots  & \dots  & \dots  & \dots  & \dots  & \dots  \\
\hline  
\hline  
\end{tabular} 
\end{small} 
\end{center}  
\end{table}

\begin{table}   
\begin{center}   
\begin{small}  
\caption{Catalog for $3,794$ Point Sources in NGC\,$3077$} 
\label{tab:n3077}   
\begin{tabular}{rrrrrrrrrrrrrrrrr}   
\\   
\hline  
\hline  
\multicolumn{1}{c}{RA} &
\multicolumn{1}{c}{Dec} &
\multicolumn{1}{c}{} &
\multicolumn{1}{c}{$m_{3.6}$} &
\multicolumn{1}{c}{$\sigma_{3.6}$} &
\multicolumn{1}{c}{$\delta_{3.6}$} &
\multicolumn{1}{c}{$m_{4.5}$} &
\multicolumn{1}{c}{$\sigma_{4.5}$} &
\multicolumn{1}{c}{$\delta_{4.5}$} &
\multicolumn{1}{c}{$m_{5.8}$} &
\multicolumn{1}{c}{$\sigma_{5.8}$} &
\multicolumn{1}{c}{$\delta_{5.8}$} &
\multicolumn{1}{c}{$m_{8.0}$} &
\multicolumn{1}{c}{$\sigma_{8.0}$} &
\multicolumn{1}{c}{$\delta_{8.0}$} &
\multicolumn{1}{c}{$m_{24}$} &
\multicolumn{1}{c}{$\sigma_{24}$} 
\\
\multicolumn{1}{c}{(deg)} &
\multicolumn{1}{c}{(deg)} &
\multicolumn{1}{c}{} &
\multicolumn{1}{c}{(mag)} &
\multicolumn{1}{c}{} &
\multicolumn{1}{c}{} &
\multicolumn{1}{c}{(mag)} &
\multicolumn{1}{c}{} &
\multicolumn{1}{c}{} &
\multicolumn{1}{c}{(mag)} &
\multicolumn{1}{c}{} &
\multicolumn{1}{c}{} &
\multicolumn{1}{c}{(mag)} &
\multicolumn{1}{c}{} &
\multicolumn{1}{c}{} &
\multicolumn{1}{c}{(mag)} &
\multicolumn{1}{c}{} 
\\ 
\hline  
\hline
\dots        & \dots       && \dots  & \dots  & \dots  & \dots  & \dots  & \dots  & \dots  & \dots  & \dots  & \dots  & \dots  & \dots  & \dots  & \dots  \\
 $150.81010$  &  $68.73050$   &&   $16.25$  &  $0.10$  &  $0.49$  &  $ 16.45$  &  $0.11$  &  $0.81$  &  $ 14.73$  &  $0.07$  &  $99.99$  &  $ 13.46$  &  $0.12$  &  $99.99$  &  $  8.40$  &  $0.02   $  \\  
 $150.53804$  &  $68.81851$   &&   $16.28$  &  $0.07$  &  $-0.04$  &  $ 16.45$  &  $0.07$  &  $0.15$  &  $ 16.52$  &  $0.14$  &  $-0.07$  &  $ 15.47$  &  $0.05$  &  $-0.57$  &  $  99.99$  &  $99.99   $  \\  
 $150.82225$  &  $68.74129$   &&   $16.49$  &  $0.09$  &  $0.30$  &  $ 16.45$  &  $0.12$  &  $99.99$  &  $ 16.21$  &  $0.13$  &  $1.27$  &  $ 13.62$  &  $0.12$  &  $99.99$  &  $  8.16$  &  $0.17   $  \\  
 $150.86320$  &  $68.69979$   &&   $17.13$  &  $0.07$  &  $0.47$  &  $ 16.45$  &  $0.06$  &  $0.17$  &  $ 16.21$  &  $0.05$  &  $-0.55$  &  $ 15.79$  &  $0.11$  &  $-0.87$  &  $ 12.87$  &  $0.20   $  \\  
 $150.99858$  &  $68.69220$   &&   $17.14$  &  $0.11$  &  $0.70$  &  $ 16.46$  &  $0.06$  &  $0.17$  &  $ 16.36$  &  $0.10$  &  $0.35$  &  $ 15.42$  &  $0.06$  &  $0.05$  &  $ 11.33$  &  $0.08 $  \\  
\dots        & \dots       && \dots  & \dots  & \dots  & \dots  & \dots  & \dots  & \dots  & \dots  & \dots  & \dots  & \dots  & \dots  & \dots  & \dots  \\
\hline  
\hline  
\end{tabular} 
\end{small} 
\end{center}  
\end{table}

\end{landscape}

\clearpage

\begin{landscape}

\begin{table}   
\begin{center}   
\begin{small}  
\caption{Catalog for $6,972$ Point Sources in NGC\,$1313$} 
\label{tab:n1313}   
\begin{tabular}{rrrrrrrrrrrrrrrrr}   
\\   
\hline  
\hline  
\multicolumn{1}{c}{RA} &
\multicolumn{1}{c}{Dec} &
\multicolumn{1}{c}{} &
\multicolumn{1}{c}{$m_{3.6}$} &
\multicolumn{1}{c}{$\sigma_{3.6}$} &
\multicolumn{1}{c}{$\delta_{3.6}$} &
\multicolumn{1}{c}{$m_{4.5}$} &
\multicolumn{1}{c}{$\sigma_{4.5}$} &
\multicolumn{1}{c}{$\delta_{4.5}$} &
\multicolumn{1}{c}{$m_{5.8}$} &
\multicolumn{1}{c}{$\sigma_{5.8}$} &
\multicolumn{1}{c}{$\delta_{5.8}$} &
\multicolumn{1}{c}{$m_{8.0}$} &
\multicolumn{1}{c}{$\sigma_{8.0}$} &
\multicolumn{1}{c}{$\delta_{8.0}$} &
\multicolumn{1}{c}{$m_{24}$} &
\multicolumn{1}{c}{$\sigma_{24}$} 
\\
\multicolumn{1}{c}{(deg)} &
\multicolumn{1}{c}{(deg)} &
\multicolumn{1}{c}{} &
\multicolumn{1}{c}{(mag)} &
\multicolumn{1}{c}{} &
\multicolumn{1}{c}{} &
\multicolumn{1}{c}{(mag)} &
\multicolumn{1}{c}{} &
\multicolumn{1}{c}{} &
\multicolumn{1}{c}{(mag)} &
\multicolumn{1}{c}{} &
\multicolumn{1}{c}{} &
\multicolumn{1}{c}{(mag)} &
\multicolumn{1}{c}{} &
\multicolumn{1}{c}{} &
\multicolumn{1}{c}{(mag)} &
\multicolumn{1}{c}{} 
\\ 
\hline  
\hline
\dots        & \dots       && \dots  & \dots  & \dots  & \dots  & \dots  & \dots  & \dots  & \dots  & \dots  & \dots  & \dots  & \dots  & \dots  & \dots  \\
 $ 49.77658$  &  $-66.41660$   &&   $15.55$  &  $0.05$  &  $0.00$  &  $ 15.60$  &  $0.06$  &  $0.01$  &  $ 15.65$  &  $0.05$  &  $0.23$  &  $ 15.80$  &  $0.04$  &  $0.11$  &  $ 12.13$  &  $0.07   $  \\  
 $ 49.56655$  &  $-66.49622$   &&   $15.60$  &  $0.12$  &  $0.13$  &  $ 15.60$  &  $0.13$  &  $-0.03$  &  $ 14.38$  &  $0.08$  &  $99.99$  &  $ 13.28$  &  $0.07$  &  $99.99$  &  $  8.47$  &  $0.29   $  \\  
 $ 49.57424$  &  $-66.47501$   &&   $16.41$  &  $0.13$  &  $0.08$  &  $ 15.60$  &  $0.02$  &  $-0.00$  &  $ 14.82$  &  $0.04$  &  $0.30$  &  $ 13.92$  &  $0.06$  &  $0.74$  &  $  8.44$  &  $0.04   $  \\  
 $ 49.55194$  &  $-66.50012$   &&   $16.43$  &  $0.06$  &  $0.18$  &  $ 15.61$  &  $0.05$  &  $0.07$  &  $ 14.91$  &  $0.02$  &  $0.16$  &  $ 13.72$  &  $0.05$  &  $0.39$  &  $ 11.58$  &  $99.99   $  \\  
 $ 49.56185$  &  $-66.49652$   &&   $15.52$  &  $0.04$  &  $0.01$  &  $ 15.61$  &  $0.05$  &  $0.10$  &  $ 14.88$  &  $0.05$  &  $0.54$  &  $ 14.04$  &  $0.08$  &  $1.28$  &  $  8.42$  &  $0.07  $  \\  
\dots        & \dots       && \dots  & \dots  & \dots  & \dots  & \dots  & \dots  & \dots  & \dots  & \dots  & \dots  & \dots  & \dots  & \dots  & \dots  \\
\hline  
\hline  
\end{tabular} 
\end{small} 
\end{center}  
\end{table}

\begin{table}   
\begin{center}   
\begin{small}  
\caption{Catalog for $10,264$ Point Sources in NGC\,$4736$ (M\,94)} 
\label{tab:n4736}   
\begin{tabular}{rrrrrrrrrrrrrrrrr}   
\\   
\hline  
\hline  
\multicolumn{1}{c}{RA} &
\multicolumn{1}{c}{Dec} &
\multicolumn{1}{c}{} &
\multicolumn{1}{c}{$m_{3.6}$} &
\multicolumn{1}{c}{$\sigma_{3.6}$} &
\multicolumn{1}{c}{$\delta_{3.6}$} &
\multicolumn{1}{c}{$m_{4.5}$} &
\multicolumn{1}{c}{$\sigma_{4.5}$} &
\multicolumn{1}{c}{$\delta_{4.5}$} &
\multicolumn{1}{c}{$m_{5.8}$} &
\multicolumn{1}{c}{$\sigma_{5.8}$} &
\multicolumn{1}{c}{$\delta_{5.8}$} &
\multicolumn{1}{c}{$m_{8.0}$} &
\multicolumn{1}{c}{$\sigma_{8.0}$} &
\multicolumn{1}{c}{$\delta_{8.0}$} &
\multicolumn{1}{c}{$m_{24}$} &
\multicolumn{1}{c}{$\sigma_{24}$} 
\\
\multicolumn{1}{c}{(deg)} &
\multicolumn{1}{c}{(deg)} &
\multicolumn{1}{c}{} &
\multicolumn{1}{c}{(mag)} &
\multicolumn{1}{c}{} &
\multicolumn{1}{c}{} &
\multicolumn{1}{c}{(mag)} &
\multicolumn{1}{c}{} &
\multicolumn{1}{c}{} &
\multicolumn{1}{c}{(mag)} &
\multicolumn{1}{c}{} &
\multicolumn{1}{c}{} &
\multicolumn{1}{c}{(mag)} &
\multicolumn{1}{c}{} &
\multicolumn{1}{c}{} &
\multicolumn{1}{c}{(mag)} &
\multicolumn{1}{c}{} 
\\ 
\hline  
\hline
\dots        & \dots       && \dots  & \dots  & \dots  & \dots  & \dots  & \dots  & \dots  & \dots  & \dots  & \dots  & \dots  & \dots  & \dots  & \dots  \\
 $192.70675$  &  $41.13108$   &&   $15.65$  &  $0.09$  &  $0.78$  &  $ 15.11$  &  $0.11$  &  $0.46$  &  $ 13.03$  &  $0.13$  &  $0.80$  &  $ 12.19$  &  $0.14$  &  $1.80$  &  $  6.20$  &  $0.01   $  \\  
 $192.80644$  &  $41.16393$   &&   $15.38$  &  $0.08$  &  $0.12$  &  $ 15.11$  &  $0.04$  &  $0.12$  &  $ 15.34$  &  $0.04$  &  $0.28$  &  $ 15.04$  &  $0.06$  &  $0.35$  &  $ 12.35$  &  $0.26   $  \\  
 $192.79610$  &  $41.14728$   &&   $16.13$  &  $0.06$  &  $0.15$  &  $ 15.11$  &  $0.03$  &  $0.05$  &  $ 14.38$  &  $0.06$  &  $-0.00$  &  $ 13.50$  &  $0.03$  &  $0.07$  &  $  9.87$  &  $0.03   $  \\  
 $192.73827$  &  $41.11505$   &&   $15.44$  &  $0.12$  &  $1.36$  &  $ 15.11$  &  $0.12$  &  $1.19$  &  $ 12.88$  &  $0.10$  &  $1.02$  &  $ 11.62$  &  $0.07$  &  $1.79$  &  $  5.57$  &  $0.06   $  \\  
 $192.70926$  &  $41.19135$   &&   $15.14$  &  $0.09$  &  $1.11$  &  $ 15.12$  &  $0.08$  &  $1.24$  &  $ 15.07$  &  $0.09$  &  $1.64$  &  $ 12.09$  &  $0.08$  &  $0.99$  &  $  8.28$  &  $0.01  $  \\  
\dots        & \dots       && \dots  & \dots  & \dots  & \dots  & \dots  & \dots  & \dots  & \dots  & \dots  & \dots  & \dots  & \dots  & \dots  & \dots  \\
\hline  
\hline  
\end{tabular} 
\end{small} 
\end{center}  
\end{table}

\end{landscape}

\clearpage

\begin{landscape}

\begin{table}   
\begin{center}   
\begin{small}  
\caption{Catalog for $5,137$ Point Sources in NGC\,$4826$(M\,$64$)} 
\label{tab:n4826}   
\begin{tabular}{rrrrrrrrrrrrrrrrr}   
\\   
\hline  
\hline  
\multicolumn{1}{c}{RA} &
\multicolumn{1}{c}{Dec} &
\multicolumn{1}{c}{} &
\multicolumn{1}{c}{$m_{3.6}$} &
\multicolumn{1}{c}{$\sigma_{3.6}$} &
\multicolumn{1}{c}{$\delta_{3.6}$} &
\multicolumn{1}{c}{$m_{4.5}$} &
\multicolumn{1}{c}{$\sigma_{4.5}$} &
\multicolumn{1}{c}{$\delta_{4.5}$} &
\multicolumn{1}{c}{$m_{5.8}$} &
\multicolumn{1}{c}{$\sigma_{5.8}$} &
\multicolumn{1}{c}{$\delta_{5.8}$} &
\multicolumn{1}{c}{$m_{8.0}$} &
\multicolumn{1}{c}{$\sigma_{8.0}$} &
\multicolumn{1}{c}{$\delta_{8.0}$} &
\multicolumn{1}{c}{$m_{24}$} &
\multicolumn{1}{c}{$\sigma_{24}$} 
\\
\multicolumn{1}{c}{(deg)} &
\multicolumn{1}{c}{(deg)} &
\multicolumn{1}{c}{} &
\multicolumn{1}{c}{(mag)} &
\multicolumn{1}{c}{} &
\multicolumn{1}{c}{} &
\multicolumn{1}{c}{(mag)} &
\multicolumn{1}{c}{} &
\multicolumn{1}{c}{} &
\multicolumn{1}{c}{(mag)} &
\multicolumn{1}{c}{} &
\multicolumn{1}{c}{} &
\multicolumn{1}{c}{(mag)} &
\multicolumn{1}{c}{} &
\multicolumn{1}{c}{} &
\multicolumn{1}{c}{(mag)} &
\multicolumn{1}{c}{} 
\\ 
\hline  
\hline
\dots        & \dots       && \dots  & \dots  & \dots  & \dots  & \dots  & \dots  & \dots  & \dots  & \dots  & \dots  & \dots  & \dots  & \dots  & \dots  \\
 $194.06033$  &  $21.73660$   &&   $15.41$  &  $0.10$  &  $0.13$  &  $ 15.34$  &  $0.08$  &  $0.07$  &  $ 15.73$  &  $0.16$  &  $0.73$  &  $ 15.02$  &  $0.11$  &  $-0.10$  &  $ 11.52$  &  $0.05   $  \\  
 $194.05749$  &  $21.67165$   &&   $16.50$  &  $0.05$  &  $0.04$  &  $ 15.35$  &  $0.10$  &  $-0.16$  &  $ 14.36$  &  $0.06$  &  $-0.04$  &  $ 12.90$  &  $0.05$  &  $-0.14$  &  $  9.68$  &  $0.03   $  \\  
 $194.21598$  &  $21.66067$   &&   $15.31$  &  $0.06$  &  $-0.14$  &  $ 15.36$  &  $0.07$  &  $-0.02$  &  $ 15.33$  &  $0.05$  &  $-0.04$  &  $ 15.40$  &  $0.08$  &  $-0.41$  &  $ 13.06$  &  $99.99   $  \\  
 $194.16860$  &  $21.68549$   &&   $15.37$  &  $0.13$  &  $99.99$  &  $ 15.37$  &  $0.13$  &  $99.99$  &  $ 13.62$  &  $0.10$  &  $0.59$  &  $ 11.61$  &  $0.09$  &  $0.49$  &  $  6.93$  &  $0.03   $  \\  
 $194.12018$  &  $21.64330$   &&   $15.51$  &  $0.07$  &  $0.33$  &  $ 15.38$  &  $0.07$  &  $0.37$  &  $ 14.95$  &  $0.10$  &  $0.26$  &  $ 14.09$  &  $0.05$  &  $-0.02$  &  $ 10.57$  &  $0.07 $  \\  
\dots        & \dots       && \dots  & \dots  & \dots  & \dots  & \dots  & \dots  & \dots  & \dots  & \dots  & \dots  & \dots  & \dots  & \dots  & \dots  \\
\hline  
\hline  
\end{tabular} 
\end{small} 
\end{center}  
\end{table}

\begin{table}   
\begin{center}   
\begin{small}  
\caption{Catalog for $23,331$ Point Sources in NGC\,$5236$ (M\,$83$)} 
\label{tab:m83}   
\begin{tabular}{rrrrrrrrrrrrrrrrr}   
\\   
\hline  
\hline  
\multicolumn{1}{c}{RA} &
\multicolumn{1}{c}{Dec} &
\multicolumn{1}{c}{} &
\multicolumn{1}{c}{$m_{3.6}$} &
\multicolumn{1}{c}{$\sigma_{3.6}$} &
\multicolumn{1}{c}{$\delta_{3.6}$} &
\multicolumn{1}{c}{$m_{4.5}$} &
\multicolumn{1}{c}{$\sigma_{4.5}$} &
\multicolumn{1}{c}{$\delta_{4.5}$} &
\multicolumn{1}{c}{$m_{5.8}$} &
\multicolumn{1}{c}{$\sigma_{5.8}$} &
\multicolumn{1}{c}{$\delta_{5.8}$} &
\multicolumn{1}{c}{$m_{8.0}$} &
\multicolumn{1}{c}{$\sigma_{8.0}$} &
\multicolumn{1}{c}{$\delta_{8.0}$} &
\multicolumn{1}{c}{$m_{24}$} &
\multicolumn{1}{c}{$\sigma_{24}$} 
\\
\multicolumn{1}{c}{(deg)} &
\multicolumn{1}{c}{(deg)} &
\multicolumn{1}{c}{} &
\multicolumn{1}{c}{(mag)} &
\multicolumn{1}{c}{} &
\multicolumn{1}{c}{} &
\multicolumn{1}{c}{(mag)} &
\multicolumn{1}{c}{} &
\multicolumn{1}{c}{} &
\multicolumn{1}{c}{(mag)} &
\multicolumn{1}{c}{} &
\multicolumn{1}{c}{} &
\multicolumn{1}{c}{(mag)} &
\multicolumn{1}{c}{} &
\multicolumn{1}{c}{} &
\multicolumn{1}{c}{(mag)} &
\multicolumn{1}{c}{} 
\\ 
\hline  
\hline
\dots        & \dots       && \dots  & \dots  & \dots  & \dots  & \dots  & \dots  & \dots  & \dots  & \dots  & \dots  & \dots  & \dots  & \dots  & \dots  \\
 $204.15718$  &  $-29.85278$   &&   $13.49$  &  $0.04$  &  $-0.03$  &  $ 13.54$  &  $0.03$  &  $-0.01$  &  $ 13.49$  &  $0.02$  &  $-0.05$  &  $ 13.57$  &  $0.05$  &  $0.25$  &  $ 11.95$  &  $99.99   $  \\  
 $204.15082$  &  $-29.74423$   &&   $13.50$  &  $0.03$  &  $-0.03$  &  $ 13.55$  &  $0.02$  &  $-0.01$  &  $ 13.52$  &  $0.03$  &  $-0.01$  &  $ 13.55$  &  $0.02$  &  $-0.05$  &  $ 11.89$  &  $99.99   $  \\  
 $204.23002$  &  $-29.97769$   &&   $13.75$  &  $0.06$  &  $0.05$  &  $ 13.55$  &  $0.03$  &  $-0.08$  &  $ 13.64$  &  $0.02$  &  $0.01$  &  $ 13.89$  &  $0.04$  &  $0.03$  &  $ 12.31$  &  $99.99   $  \\  
 $204.22073$  &  $-29.86514$   &&   $14.09$  &  $0.09$  &  $0.93$  &  $ 13.55$  &  $0.08$  &  $0.75$  &  $ 10.94$  &  $0.05$  &  $0.77$  &  $  8.38$  &  $0.03$  &  $99.99$  &  $  4.18$  &  $0.03   $  \\  
 $204.18453$  &  $-29.87696$   &&   $13.57$  &  $0.04$  &  $-0.01$  &  $ 13.56$  &  $0.04$  &  $-0.05$  &  $ 13.15$  &  $0.05$  &  $0.18$  &  $ 12.74$  &  $0.09$  &  $0.95$  &  $  9.07$  &  $99.99  $  \\  
\dots        & \dots       && \dots  & \dots  & \dots  & \dots  & \dots  & \dots  & \dots  & \dots  & \dots  & \dots  & \dots  & \dots  & \dots  & \dots  \\
\hline  
\hline  
\end{tabular} 
\end{small} 
\end{center}  
\end{table}

\end{landscape}

\clearpage

\begin{landscape}

\begin{table}   
\begin{center}   
\begin{small}  
\caption{Catalog for $5,617$ Point Sources in NGC\,$5068$} 
\label{tab:n5068}   
\begin{tabular}{rrrrrrrrrrrrrrrrr}   
\\   
\hline  
\hline  
\multicolumn{1}{c}{RA} &
\multicolumn{1}{c}{Dec} &
\multicolumn{1}{c}{} &
\multicolumn{1}{c}{$m_{3.6}$} &
\multicolumn{1}{c}{$\sigma_{3.6}$} &
\multicolumn{1}{c}{$\delta_{3.6}$} &
\multicolumn{1}{c}{$m_{4.5}$} &
\multicolumn{1}{c}{$\sigma_{4.5}$} &
\multicolumn{1}{c}{$\delta_{4.5}$} &
\multicolumn{1}{c}{$m_{5.8}$} &
\multicolumn{1}{c}{$\sigma_{5.8}$} &
\multicolumn{1}{c}{$\delta_{5.8}$} &
\multicolumn{1}{c}{$m_{8.0}$} &
\multicolumn{1}{c}{$\sigma_{8.0}$} &
\multicolumn{1}{c}{$\delta_{8.0}$} &
\multicolumn{1}{c}{$m_{24}$} &
\multicolumn{1}{c}{$\sigma_{24}$} 
\\
\multicolumn{1}{c}{(deg)} &
\multicolumn{1}{c}{(deg)} &
\multicolumn{1}{c}{} &
\multicolumn{1}{c}{(mag)} &
\multicolumn{1}{c}{} &
\multicolumn{1}{c}{} &
\multicolumn{1}{c}{(mag)} &
\multicolumn{1}{c}{} &
\multicolumn{1}{c}{} &
\multicolumn{1}{c}{(mag)} &
\multicolumn{1}{c}{} &
\multicolumn{1}{c}{} &
\multicolumn{1}{c}{(mag)} &
\multicolumn{1}{c}{} &
\multicolumn{1}{c}{} &
\multicolumn{1}{c}{(mag)} &
\multicolumn{1}{c}{} 
\\ 
\hline  
\hline
\dots        & \dots       && \dots  & \dots  & \dots  & \dots  & \dots  & \dots  & \dots  & \dots  & \dots  & \dots  & \dots  & \dots  & \dots  & \dots  \\
 $199.72607$  &  $-21.03384$   &&   $15.92$  &  $0.14$  &  $0.83$  &  $ 15.41$  &  $0.11$  &  $0.37$  &  $ 13.75$  &  $0.11$  &  $0.71$  &  $ 11.39$  &  $0.04$  &  $0.23$  &  $  7.91$  &  $0.07   $  \\  
 $199.73146$  &  $-21.04449$   &&   $15.61$  &  $0.13$  &  $0.05$  &  $ 15.41$  &  $0.09$  &  $-0.10$  &  $ 13.50$  &  $0.09$  &  $-0.03$  &  $ 11.48$  &  $0.02$  &  $-0.04$  &  $  7.87$  &  $0.11   $  \\  
 $199.70099$  &  $-20.97649$   &&   $15.48$  &  $0.13$  &  $0.88$  &  $ 15.42$  &  $0.13$  &  $0.85$  &  $ 14.22$  &  $0.11$  &  $0.88$  &  $ 11.91$  &  $0.06$  &  $0.64$  &  $  8.32$  &  $0.02   $  \\  
 $199.73044$  &  $-21.04431$   &&   $15.22$  &  $0.14$  &  $0.23$  &  $ 15.43$  &  $0.12$  &  $0.40$  &  $ 13.26$  &  $0.11$  &  $0.08$  &  $ 11.00$  &  $0.04$  &  $-0.38$  &  $  7.89$  &  $0.05   $  \\  
 $199.71533$  &  $-21.07158$   &&   $15.37$  &  $0.05$  &  $0.02$  &  $ 15.43$  &  $0.04$  &  $0.02$  &  $ 15.08$  &  $0.07$  &  $-0.38$  &  $ 13.84$  &  $0.07$  &  $0.37$  &  $  9.70$  &  $0.05   $  \\  
\dots        & \dots       && \dots  & \dots  & \dots  & \dots  & \dots  & \dots  & \dots  & \dots  & \dots  & \dots  & \dots  & \dots  & \dots  & \dots  \\
\hline  
\hline  
\end{tabular} 
\end{small} 
\end{center}  
\end{table}

\begin{table}   
\begin{center}   
\begin{small}  
\caption{Catalog for $15,813$ Point Sources in NGC\,$6946$} 
\label{tab:n6946}   
\begin{tabular}{rrrrrrrrrrrrrrrrr}   
\\   
\hline  
\hline  
\multicolumn{1}{c}{RA} &
\multicolumn{1}{c}{Dec} &
\multicolumn{1}{c}{} &
\multicolumn{1}{c}{$m_{3.6}$} &
\multicolumn{1}{c}{$\sigma_{3.6}$} &
\multicolumn{1}{c}{$\delta_{3.6}$} &
\multicolumn{1}{c}{$m_{4.5}$} &
\multicolumn{1}{c}{$\sigma_{4.5}$} &
\multicolumn{1}{c}{$\delta_{4.5}$} &
\multicolumn{1}{c}{$m_{5.8}$} &
\multicolumn{1}{c}{$\sigma_{5.8}$} &
\multicolumn{1}{c}{$\delta_{5.8}$} &
\multicolumn{1}{c}{$m_{8.0}$} &
\multicolumn{1}{c}{$\sigma_{8.0}$} &
\multicolumn{1}{c}{$\delta_{8.0}$} &
\multicolumn{1}{c}{$m_{24}$} &
\multicolumn{1}{c}{$\sigma_{24}$} 
\\
\multicolumn{1}{c}{(deg)} &
\multicolumn{1}{c}{(deg)} &
\multicolumn{1}{c}{} &
\multicolumn{1}{c}{(mag)} &
\multicolumn{1}{c}{} &
\multicolumn{1}{c}{} &
\multicolumn{1}{c}{(mag)} &
\multicolumn{1}{c}{} &
\multicolumn{1}{c}{} &
\multicolumn{1}{c}{(mag)} &
\multicolumn{1}{c}{} &
\multicolumn{1}{c}{} &
\multicolumn{1}{c}{(mag)} &
\multicolumn{1}{c}{} &
\multicolumn{1}{c}{} &
\multicolumn{1}{c}{(mag)} &
\multicolumn{1}{c}{} 
\\ 
\hline  
\hline
\dots        & \dots       && \dots  & \dots  & \dots  & \dots  & \dots  & \dots  & \dots  & \dots  & \dots  & \dots  & \dots  & \dots  & \dots  & \dots  \\
 $308.65553$  &  $60.13594$   &&   $14.78$  &  $0.06$  &  $-0.11$  &  $ 14.59$  &  $0.06$  &  $-0.30$  &  $ 14.40$  &  $0.04$  &  $99.99$  &  $ 13.56$  &  $0.24$  &  $99.99$  &  $  8.21$  &  $0.26   $  \\  
 $308.94799$  &  $59.96213$   &&   $14.44$  &  $0.16$  &  $-0.02$  &  $ 14.60$  &  $0.14$  &  $0.13$  &  $ 14.36$  &  $0.09$  &  $-0.02$  &  $ 14.42$  &  $0.15$  &  $0.06$  &  $  99.99$  &  $99.99   $  \\  
 $308.64470$  &  $60.09664$   &&   $14.57$  &  $0.04$  &  $-0.06$  &  $ 14.60$  &  $0.04$  &  $0.00$  &  $ 14.38$  &  $0.05$  &  $-0.03$  &  $ 14.20$  &  $0.04$  &  $-0.23$  &  $ 11.33$  &  $99.99   $  \\  
 $308.81946$  &  $60.18289$   &&   $14.80$  &  $0.14$  &  $0.93$  &  $ 14.60$  &  $0.10$  &  $1.15$  &  $ 12.19$  &  $0.10$  &  $1.13$  &  $ 11.01$  &  $0.12$  &  $1.71$  &  $  5.14$  &  $0.01   $  \\  
 $309.03955$  &  $60.09698$   &&   $13.45$  &  $0.11$  &  $0.56$  &  $ 14.60$  &  $0.10$  &  $0.67$  &  $ 12.25$  &  $0.12$  &  $0.59$  &  $ 12.32$  &  $99.99$  &  $99.99$  &  $  99.99$  &  $99.99 $  \\  
\dots        & \dots       && \dots  & \dots  & \dots  & \dots  & \dots  & \dots  & \dots  & \dots  & \dots  & \dots  & \dots  & \dots  & \dots  & \dots  \\
\hline  
\hline  
\end{tabular} 
\end{small} 
\end{center}  
\end{table}

\end{landscape}

\clearpage

\begin{landscape}

\begin{table}   
\begin{center}   
\begin{small}  
\caption{Catalog for $991$ Point Sources in NGC\,$5474$} 
\label{tab:n5474}   
\begin{tabular}{rrrrrrrrrrrrrrrrr}   
\\   
\hline  
\hline  
\multicolumn{1}{c}{RA} &
\multicolumn{1}{c}{Dec} &
\multicolumn{1}{c}{} &
\multicolumn{1}{c}{$m_{3.6}$} &
\multicolumn{1}{c}{$\sigma_{3.6}$} &
\multicolumn{1}{c}{$\delta_{3.6}$} &
\multicolumn{1}{c}{$m_{4.5}$} &
\multicolumn{1}{c}{$\sigma_{4.5}$} &
\multicolumn{1}{c}{$\delta_{4.5}$} &
\multicolumn{1}{c}{$m_{5.8}$} &
\multicolumn{1}{c}{$\sigma_{5.8}$} &
\multicolumn{1}{c}{$\delta_{5.8}$} &
\multicolumn{1}{c}{$m_{8.0}$} &
\multicolumn{1}{c}{$\sigma_{8.0}$} &
\multicolumn{1}{c}{$\delta_{8.0}$} &
\multicolumn{1}{c}{$m_{24}$} &
\multicolumn{1}{c}{$\sigma_{24}$} 
\\
\multicolumn{1}{c}{(deg)} &
\multicolumn{1}{c}{(deg)} &
\multicolumn{1}{c}{} &
\multicolumn{1}{c}{(mag)} &
\multicolumn{1}{c}{} &
\multicolumn{1}{c}{} &
\multicolumn{1}{c}{(mag)} &
\multicolumn{1}{c}{} &
\multicolumn{1}{c}{} &
\multicolumn{1}{c}{(mag)} &
\multicolumn{1}{c}{} &
\multicolumn{1}{c}{} &
\multicolumn{1}{c}{(mag)} &
\multicolumn{1}{c}{} &
\multicolumn{1}{c}{} &
\multicolumn{1}{c}{(mag)} &
\multicolumn{1}{c}{} 
\\ 
\hline  
\hline
\dots        & \dots       && \dots  & \dots  & \dots  & \dots  & \dots  & \dots  & \dots  & \dots  & \dots  & \dots  & \dots  & \dots  & \dots  & \dots  \\
 $211.20194$  &  $53.64062$   &&   $18.02$  &  $0.05$  &  $0.06$  &  $ 17.17$  &  $0.07$  &  $-0.19$  &  $ 16.70$  &  $0.11$  &  $99.99$  &  $ 16.18$  &  $0.12$  &  $0.49$  &  $ 11.28$  &  $0.05   $  \\  
 $211.29572$  &  $53.67260$   &&   $17.25$  &  $0.06$  &  $0.02$  &  $ 17.17$  &  $0.06$  &  $0.06$  &  $ 16.88$  &  $0.08$  &  $-0.09$  &  $ 15.81$  &  $0.12$  &  $-0.43$  &  $ 11.49$  &  $0.05   $  \\  
 $211.25098$  &  $53.63765$   &&   $18.05$  &  $0.16$  &  $1.14$  &  $ 17.18$  &  $0.13$  &  $0.80$  &  $ 16.08$  &  $0.09$  &  $0.49$  &  $ 14.49$  &  $0.07$  &  $0.42$  &  $ 10.31$  &  $0.06   $  \\  
 $211.23574$  &  $53.66682$   &&   $17.37$  &  $0.05$  &  $-0.07$  &  $ 17.19$  &  $0.04$  &  $-0.02$  &  $ 16.71$  &  $0.17$  &  $-0.32$  &  $ 15.94$  &  $0.11$  &  $0.80$  &  $ 11.06$  &  $0.09   $  \\  
 $211.29303$  &  $53.71026$   &&   $17.28$  &  $0.08$  &  $-0.03$  &  $ 17.21$  &  $0.13$  &  $0.45$  &  $ 17.16$  &  $0.15$  &  $-0.25$  &  $ 16.12$  &  $0.23$  &  $0.22$  &  $ 12.94$  &  $0.27 $  \\  
\dots        & \dots       && \dots  & \dots  & \dots  & \dots  & \dots  & \dots  & \dots  & \dots  & \dots  & \dots  & \dots  & \dots  & \dots  & \dots  \\
\hline  
\hline  
\end{tabular} 
\end{small} 
\end{center}  
\end{table}

\begin{table}   
\begin{center}   
\begin{small}  
\caption{Catalog for $16,291$ Point Sources in NGC\,$5457$ (M\,$101$)} 
\label{tab:m101}   
\begin{tabular}{rrrrrrrrrrrrrrrrr}   
\\   
\hline  
\hline  
\multicolumn{1}{c}{RA} &
\multicolumn{1}{c}{Dec} &
\multicolumn{1}{c}{} &
\multicolumn{1}{c}{$m_{3.6}$} &
\multicolumn{1}{c}{$\sigma_{3.6}$} &
\multicolumn{1}{c}{$\delta_{3.6}$} &
\multicolumn{1}{c}{$m_{4.5}$} &
\multicolumn{1}{c}{$\sigma_{4.5}$} &
\multicolumn{1}{c}{$\delta_{4.5}$} &
\multicolumn{1}{c}{$m_{5.8}$} &
\multicolumn{1}{c}{$\sigma_{5.8}$} &
\multicolumn{1}{c}{$\delta_{5.8}$} &
\multicolumn{1}{c}{$m_{8.0}$} &
\multicolumn{1}{c}{$\sigma_{8.0}$} &
\multicolumn{1}{c}{$\delta_{8.0}$} &
\multicolumn{1}{c}{$m_{24}$} &
\multicolumn{1}{c}{$\sigma_{24}$} 
\\
\multicolumn{1}{c}{(deg)} &
\multicolumn{1}{c}{(deg)} &
\multicolumn{1}{c}{} &
\multicolumn{1}{c}{(mag)} &
\multicolumn{1}{c}{} &
\multicolumn{1}{c}{} &
\multicolumn{1}{c}{(mag)} &
\multicolumn{1}{c}{} &
\multicolumn{1}{c}{} &
\multicolumn{1}{c}{(mag)} &
\multicolumn{1}{c}{} &
\multicolumn{1}{c}{} &
\multicolumn{1}{c}{(mag)} &
\multicolumn{1}{c}{} &
\multicolumn{1}{c}{} &
\multicolumn{1}{c}{(mag)} &
\multicolumn{1}{c}{} 
\\ 
\hline  
\hline
\dots        & \dots       && \dots  & \dots  & \dots  & \dots  & \dots  & \dots  & \dots  & \dots  & \dots  & \dots  & \dots  & \dots  & \dots  & \dots  \\
 $210.78059$  &  $54.33504$   &&   $14.36$  &  $0.09$  &  $0.01$  &  $ 14.33$  &  $0.07$  &  $0.01$  &  $ 14.38$  &  $0.03$  &  $-0.43$  &  $ 14.72$  &  $0.09$  &  $99.99$  &  $ 11.05$  &  $99.99   $  \\  
 $210.67501$  &  $54.19461$   &&   $14.28$  &  $0.05$  &  $0.00$  &  $ 14.34$  &  $0.04$  &  $0.01$  &  $ 14.55$  &  $0.05$  &  $0.10$  &  $ 14.25$  &  $0.05$  &  $0.06$  &  $ 11.79$  &  $0.19   $  \\  
 $210.77003$  &  $54.32386$   &&   $14.48$  &  $0.10$  &  $0.62$  &  $ 14.35$  &  $0.14$  &  $0.90$  &  $ 11.57$  &  $0.06$  &  $0.57$  &  $  9.73$  &  $0.02$  &  $0.51$  &  $  5.45$  &  $0.01   $  \\  
 $210.86363$  &  $54.31284$   &&   $14.77$  &  $0.09$  &  $0.43$  &  $ 14.35$  &  $0.06$  &  $0.38$  &  $ 12.01$  &  $0.07$  &  $0.51$  &  $ 10.21$  &  $0.03$  &  $0.58$  &  $  5.41$  &  $0.01   $  \\  
 $210.58743$  &  $54.46138$   &&   $14.12$  &  $0.07$  &  $-0.07$  &  $ 14.35$  &  $0.06$  &  $0.12$  &  $ 14.33$  &  $0.06$  &  $0.07$  &  $ 14.15$  &  $0.03$  &  $-0.06$  &  $ 12.01$  &  $0.29 $  \\  
\dots        & \dots       && \dots  & \dots  & \dots  & \dots  & \dots  & \dots  & \dots  & \dots  & \dots  & \dots  & \dots  & \dots  & \dots  & \dots  \\
\hline  
\hline  
\end{tabular} 
\end{small} 
\end{center}  
\end{table}

\end{landscape}

\clearpage

\begin{landscape}

\begin{table}   
\begin{center}   
\begin{small}  
\caption{Catalog for $4,321$ Point Sources in NGC\,$45$} 
\label{tab:n0045}   
\begin{tabular}{rrrrrrrrrrrrrrrrr}   
\\   
\hline  
\hline  
\multicolumn{1}{c}{RA} &
\multicolumn{1}{c}{Dec} &
\multicolumn{1}{c}{} &
\multicolumn{1}{c}{$m_{3.6}$} &
\multicolumn{1}{c}{$\sigma_{3.6}$} &
\multicolumn{1}{c}{$\delta_{3.6}$} &
\multicolumn{1}{c}{$m_{4.5}$} &
\multicolumn{1}{c}{$\sigma_{4.5}$} &
\multicolumn{1}{c}{$\delta_{4.5}$} &
\multicolumn{1}{c}{$m_{5.8}$} &
\multicolumn{1}{c}{$\sigma_{5.8}$} &
\multicolumn{1}{c}{$\delta_{5.8}$} &
\multicolumn{1}{c}{$m_{8.0}$} &
\multicolumn{1}{c}{$\sigma_{8.0}$} &
\multicolumn{1}{c}{$\delta_{8.0}$} &
\multicolumn{1}{c}{$m_{24}$} &
\multicolumn{1}{c}{$\sigma_{24}$} 
\\
\multicolumn{1}{c}{(deg)} &
\multicolumn{1}{c}{(deg)} &
\multicolumn{1}{c}{} &
\multicolumn{1}{c}{(mag)} &
\multicolumn{1}{c}{} &
\multicolumn{1}{c}{} &
\multicolumn{1}{c}{(mag)} &
\multicolumn{1}{c}{} &
\multicolumn{1}{c}{} &
\multicolumn{1}{c}{(mag)} &
\multicolumn{1}{c}{} &
\multicolumn{1}{c}{} &
\multicolumn{1}{c}{(mag)} &
\multicolumn{1}{c}{} &
\multicolumn{1}{c}{} &
\multicolumn{1}{c}{(mag)} &
\multicolumn{1}{c}{} 
\\ 
\hline  
\hline
\dots        & \dots       && \dots  & \dots  & \dots  & \dots  & \dots  & \dots  & \dots  & \dots  & \dots  & \dots  & \dots  & \dots  & \dots  & \dots  \\
 $  3.63601$  &  $-23.18667$   &&   $15.84$  &  $0.06$  &  $0.31$  &  $ 15.96$  &  $0.09$  &  $0.77$  &  $ 15.71$  &  $0.17$  &  $0.61$  &  $ 14.63$  &  $0.05$  &  $0.76$  &  $ 10.74$  &  $0.09   $  \\  
 $  3.40049$  &  $-23.19893$   &&   $16.25$  &  $0.07$  &  $0.10$  &  $ 15.96$  &  $0.13$  &  $0.39$  &  $ 15.29$  &  $0.07$  &  $-0.18$  &  $ 12.62$  &  $0.03$  &  $0.13$  &  $  9.38$  &  $0.03   $  \\  
 $  3.48302$  &  $-23.23012$   &&   $16.11$  &  $0.08$  &  $0.22$  &  $ 15.96$  &  $0.05$  &  $0.35$  &  $ 16.26$  &  $0.07$  &  $0.68$  &  $ 15.41$  &  $0.06$  &  $0.70$  &  $ 11.39$  &  $0.13   $  \\  
 $  3.54977$  &  $-23.14415$   &&   $16.34$  &  $0.07$  &  $0.19$  &  $ 15.97$  &  $0.09$  &  $0.10$  &  $ 16.12$  &  $0.09$  &  $0.09$  &  $ 15.33$  &  $0.04$  &  $-0.22$  &  $ 11.63$  &  $0.18   $  \\  
 $  3.66439$  &  $-23.20038$   &&   $16.05$  &  $0.04$  &  $0.10$  &  $ 15.97$  &  $0.09$  &  $-0.00$  &  $ 16.33$  &  $0.11$  &  $0.53$  &  $ 15.10$  &  $0.09$  &  $99.99$  &  $ 11.59$  &  $99.99  $  \\  
\dots        & \dots       && \dots  & \dots  & \dots  & \dots  & \dots  & \dots  & \dots  & \dots  & \dots  & \dots  & \dots  & \dots  & \dots  & \dots  \\
\hline  
\hline  
\end{tabular} 
\end{small} 
\end{center}  
\end{table}

\begin{table}   
\begin{center}   
\begin{small}  
\caption{Catalog for $8,601$ Point Sources in NGC\,$5194$ (M\,$51$)} 
\label{tab:m51}   
\begin{tabular}{rrrrrrrrrrrrrrrrr}   
\\   
\hline  
\hline  
\multicolumn{1}{c}{RA} &
\multicolumn{1}{c}{Dec} &
\multicolumn{1}{c}{} &
\multicolumn{1}{c}{$m_{3.6}$} &
\multicolumn{1}{c}{$\sigma_{3.6}$} &
\multicolumn{1}{c}{$\delta_{3.6}$} &
\multicolumn{1}{c}{$m_{4.5}$} &
\multicolumn{1}{c}{$\sigma_{4.5}$} &
\multicolumn{1}{c}{$\delta_{4.5}$} &
\multicolumn{1}{c}{$m_{5.8}$} &
\multicolumn{1}{c}{$\sigma_{5.8}$} &
\multicolumn{1}{c}{$\delta_{5.8}$} &
\multicolumn{1}{c}{$m_{8.0}$} &
\multicolumn{1}{c}{$\sigma_{8.0}$} &
\multicolumn{1}{c}{$\delta_{8.0}$} &
\multicolumn{1}{c}{$m_{24}$} &
\multicolumn{1}{c}{$\sigma_{24}$} 
\\
\multicolumn{1}{c}{(deg)} &
\multicolumn{1}{c}{(deg)} &
\multicolumn{1}{c}{} &
\multicolumn{1}{c}{(mag)} &
\multicolumn{1}{c}{} &
\multicolumn{1}{c}{} &
\multicolumn{1}{c}{(mag)} &
\multicolumn{1}{c}{} &
\multicolumn{1}{c}{} &
\multicolumn{1}{c}{(mag)} &
\multicolumn{1}{c}{} &
\multicolumn{1}{c}{} &
\multicolumn{1}{c}{(mag)} &
\multicolumn{1}{c}{} &
\multicolumn{1}{c}{} &
\multicolumn{1}{c}{(mag)} &
\multicolumn{1}{c}{} 
\\ 
\hline  
\hline
\dots        & \dots       && \dots  & \dots  & \dots  & \dots  & \dots  & \dots  & \dots  & \dots  & \dots  & \dots  & \dots  & \dots  & \dots  & \dots  \\
 $202.52413$  &  $47.26076$   &&   $14.32$  &  $0.11$  &  $0.68$  &  $ 13.58$  &  $0.06$  &  $-0.01$  &  $ 13.43$  &  $0.08$  &  $0.02$  &  $ 13.61$  &  $0.05$  &  $0.50$  &  $ 11.08$  &  $0.19   $  \\  
 $202.46449$  &  $47.19562$   &&   $13.53$  &  $0.14$  &  $0.59$  &  $ 13.58$  &  $0.11$  &  $-0.09$  &  $ 12.37$  &  $0.09$  &  $-0.19$  &  $ 11.49$  &  $0.07$  &  $0.96$  &  $  6.96$  &  $0.17   $  \\  
 $202.48110$  &  $47.19441$   &&   $14.27$  &  $0.14$  &  $-1.19$  &  $ 13.59$  &  $0.08$  &  $-0.82$  &  $ 11.41$  &  $0.09$  &  $0.04$  &  $  9.61$  &  $0.08$  &  $0.11$  &  $  5.43$  &  $0.17   $  \\  
 $202.48202$  &  $47.19663$   &&   $14.19$  &  $0.09$  &  $0.39$  &  $ 13.59$  &  $0.05$  &  $0.04$  &  $ 13.08$  &  $0.11$  &  $1.11$  &  $ 10.32$  &  $0.11$  &  $-0.13$  &  $  4.76$  &  $0.06   $  \\  
 $202.46688$  &  $47.21179$   &&   $14.03$  &  $0.13$  &  $0.57$  &  $ 13.64$  &  $0.06$  &  $0.66$  &  $ 11.39$  &  $0.11$  &  $0.71$  &  $  9.82$  &  $0.06$  &  $0.96$  &  $  4.65$  &  $0.03 $  \\  
\dots        & \dots       && \dots  & \dots  & \dots  & \dots  & \dots  & \dots  & \dots  & \dots  & \dots  & \dots  & \dots  & \dots  & \dots  & \dots  \\
\hline  
\hline  
\end{tabular} 
\end{small} 
\end{center}  
\end{table}

\end{landscape}

\clearpage

\begin{landscape}

\begin{table}   
\begin{center}   
\begin{small}  
\caption{Catalog for $5,579$ Point Sources in NGC\,$2903$} 
\label{tab:n2903}   
\begin{tabular}{rrrrrrrrrrrrrrrrr}   
\\   
\hline  
\hline  
\multicolumn{1}{c}{RA} &
\multicolumn{1}{c}{Dec} &
\multicolumn{1}{c}{} &
\multicolumn{1}{c}{$m_{3.6}$} &
\multicolumn{1}{c}{$\sigma_{3.6}$} &
\multicolumn{1}{c}{$\delta_{3.6}$} &
\multicolumn{1}{c}{$m_{4.5}$} &
\multicolumn{1}{c}{$\sigma_{4.5}$} &
\multicolumn{1}{c}{$\delta_{4.5}$} &
\multicolumn{1}{c}{$m_{5.8}$} &
\multicolumn{1}{c}{$\sigma_{5.8}$} &
\multicolumn{1}{c}{$\delta_{5.8}$} &
\multicolumn{1}{c}{$m_{8.0}$} &
\multicolumn{1}{c}{$\sigma_{8.0}$} &
\multicolumn{1}{c}{$\delta_{8.0}$} &
\multicolumn{1}{c}{$m_{24}$} &
\multicolumn{1}{c}{$\sigma_{24}$} 
\\
\multicolumn{1}{c}{(deg)} &
\multicolumn{1}{c}{(deg)} &
\multicolumn{1}{c}{} &
\multicolumn{1}{c}{(mag)} &
\multicolumn{1}{c}{} &
\multicolumn{1}{c}{} &
\multicolumn{1}{c}{(mag)} &
\multicolumn{1}{c}{} &
\multicolumn{1}{c}{} &
\multicolumn{1}{c}{(mag)} &
\multicolumn{1}{c}{} &
\multicolumn{1}{c}{} &
\multicolumn{1}{c}{(mag)} &
\multicolumn{1}{c}{} &
\multicolumn{1}{c}{} &
\multicolumn{1}{c}{(mag)} &
\multicolumn{1}{c}{} 
\\ 
\hline  
\hline
\dots        & \dots       && \dots  & \dots  & \dots  & \dots  & \dots  & \dots  & \dots  & \dots  & \dots  & \dots  & \dots  & \dots  & \dots  & \dots  \\
 $143.03345$  &  $21.47890$   &&   $14.79$  &  $0.09$  &  $0.78$  &  $ 14.53$  &  $0.12$  &  $0.56$  &  $ 11.80$  &  $0.10$  &  $0.46$  &  $  9.94$  &  $0.03$  &  $0.45$  &  $  6.05$  &  $0.06   $  \\  
 $142.92413$  &  $21.63127$   &&   $14.47$  &  $0.06$  &  $-0.00$  &  $ 14.53$  &  $0.06$  &  $0.03$  &  $ 14.61$  &  $0.07$  &  $-0.01$  &  $ 14.41$  &  $0.08$  &  $-0.35$  &  $ 11.75$  &  $0.25   $  \\  
 $143.04889$  &  $21.51972$   &&   $14.85$  &  $0.11$  &  $0.34$  &  $ 14.53$  &  $0.13$  &  $0.11$  &  $ 12.76$  &  $0.10$  &  $0.10$  &  $ 10.68$  &  $0.06$  &  $0.04$  &  $  7.46$  &  $0.11   $  \\  
 $142.93511$  &  $21.64432$   &&   $14.49$  &  $0.04$  &  $-0.02$  &  $ 14.54$  &  $0.11$  &  $-0.02$  &  $ 14.80$  &  $0.09$  &  $0.01$  &  $ 14.12$  &  $0.04$  &  $-0.45$  &  $ 12.26$  &  $99.99   $  \\  
 $142.84140$  &  $21.53222$   &&   $14.60$  &  $0.10$  &  $99.99$  &  $ 14.54$  &  $0.12$  &  $99.99$  &  $ 15.95$  &  $0.13$  &  $0.67$  &  $ 14.35$  &  $0.05$  &  $99.99$  &  $ 11.30$  &  $99.99 $  \\  
\dots        & \dots       && \dots  & \dots  & \dots  & \dots  & \dots  & \dots  & \dots  & \dots  & \dots  & \dots  & \dots  & \dots  & \dots  & \dots  \\
\hline  
\hline  
\end{tabular} 
\end{small} 
\end{center}  
\end{table}

\begin{table}   
\begin{center}   
\begin{small}  
\caption{Catalog for $4,217$ Point Sources in NGC\,$925$} 
\label{tab:n925}   
\begin{tabular}{rrrrrrrrrrrrrrrrr}   
\\   
\hline  
\hline  
\multicolumn{1}{c}{RA} &
\multicolumn{1}{c}{Dec} &
\multicolumn{1}{c}{} &
\multicolumn{1}{c}{$m_{3.6}$} &
\multicolumn{1}{c}{$\sigma_{3.6}$} &
\multicolumn{1}{c}{$\delta_{3.6}$} &
\multicolumn{1}{c}{$m_{4.5}$} &
\multicolumn{1}{c}{$\sigma_{4.5}$} &
\multicolumn{1}{c}{$\delta_{4.5}$} &
\multicolumn{1}{c}{$m_{5.8}$} &
\multicolumn{1}{c}{$\sigma_{5.8}$} &
\multicolumn{1}{c}{$\delta_{5.8}$} &
\multicolumn{1}{c}{$m_{8.0}$} &
\multicolumn{1}{c}{$\sigma_{8.0}$} &
\multicolumn{1}{c}{$\delta_{8.0}$} &
\multicolumn{1}{c}{$m_{24}$} &
\multicolumn{1}{c}{$\sigma_{24}$} 
\\
\multicolumn{1}{c}{(deg)} &
\multicolumn{1}{c}{(deg)} &
\multicolumn{1}{c}{} &
\multicolumn{1}{c}{(mag)} &
\multicolumn{1}{c}{} &
\multicolumn{1}{c}{} &
\multicolumn{1}{c}{(mag)} &
\multicolumn{1}{c}{} &
\multicolumn{1}{c}{} &
\multicolumn{1}{c}{(mag)} &
\multicolumn{1}{c}{} &
\multicolumn{1}{c}{} &
\multicolumn{1}{c}{(mag)} &
\multicolumn{1}{c}{} &
\multicolumn{1}{c}{} &
\multicolumn{1}{c}{(mag)} &
\multicolumn{1}{c}{} 
\\ 
\hline  
\hline
\dots        & \dots       && \dots  & \dots  & \dots  & \dots  & \dots  & \dots  & \dots  & \dots  & \dots  & \dots  & \dots  & \dots  & \dots  & \dots  \\
 $ 36.82781$  &  $33.63954$   &&   $15.73$  &  $0.07$  &  $-0.01$  &  $ 15.67$  &  $0.10$  &  $-0.01$  &  $ 15.66$  &  $0.06$  &  $-0.25$  &  $ 15.39$  &  $0.07$  &  $-0.85$  &  $ 11.90$  &  $0.28   $  \\  
 $ 36.84670$  &  $33.52045$   &&   $15.99$  &  $0.09$  &  $1.12$  &  $ 15.67$  &  $0.12$  &  $1.00$  &  $ 15.64$  &  $0.07$  &  $0.93$  &  $ 14.38$  &  $0.06$  &  $1.40$  &  $ 10.47$  &  $0.07   $  \\  
 $ 36.82457$  &  $33.57855$   &&   $15.79$  &  $0.11$  &  $0.28$  &  $ 15.68$  &  $0.13$  &  $0.48$  &  $ 14.56$  &  $0.14$  &  $1.16$  &  $ 12.45$  &  $0.10$  &  $0.72$  &  $  7.79$  &  $0.10   $  \\  
 $ 36.92907$  &  $33.68388$   &&   $15.63$  &  $0.06$  &  $0.00$  &  $ 15.68$  &  $0.09$  &  $0.00$  &  $ 15.94$  &  $0.11$  &  $-0.42$  &  $ 15.12$  &  $0.14$  &  $-0.56$  &  $  99.99$  &  $99.99   $  \\  
 $ 36.93849$  &  $33.57978$   &&   $15.93$  &  $0.05$  &  $0.15$  &  $ 15.68$  &  $0.09$  &  $0.10$  &  $ 15.61$  &  $0.09$  &  $-0.01$  &  $ 14.26$  &  $0.05$  &  $-0.02$  &  $ 11.85$  &  $0.12 $  \\  
\dots        & \dots       && \dots  & \dots  & \dots  & \dots  & \dots  & \dots  & \dots  & \dots  & \dots  & \dots  & \dots  & \dots  & \dots  & \dots  \\
\hline  
\hline  
\end{tabular} 
\end{small} 
\end{center}  
\end{table}

\end{landscape}

\clearpage

\begin{landscape}

\begin{table}   
\begin{center}   
\begin{small}  
\caption{Catalog for $3,102$ Point Sources in NGC\,$3627$} 
\label{tab:n3102}   
\begin{tabular}{rrrrrrrrrrrrrrrrr}   
\\   
\hline  
\hline  
\multicolumn{1}{c}{RA} &
\multicolumn{1}{c}{Dec} &
\multicolumn{1}{c}{} &
\multicolumn{1}{c}{$m_{3.6}$} &
\multicolumn{1}{c}{$\sigma_{3.6}$} &
\multicolumn{1}{c}{$\delta_{3.6}$} &
\multicolumn{1}{c}{$m_{4.5}$} &
\multicolumn{1}{c}{$\sigma_{4.5}$} &
\multicolumn{1}{c}{$\delta_{4.5}$} &
\multicolumn{1}{c}{$m_{5.8}$} &
\multicolumn{1}{c}{$\sigma_{5.8}$} &
\multicolumn{1}{c}{$\delta_{5.8}$} &
\multicolumn{1}{c}{$m_{8.0}$} &
\multicolumn{1}{c}{$\sigma_{8.0}$} &
\multicolumn{1}{c}{$\delta_{8.0}$} &
\multicolumn{1}{c}{$m_{24}$} &
\multicolumn{1}{c}{$\sigma_{24}$} 
\\
\multicolumn{1}{c}{(deg)} &
\multicolumn{1}{c}{(deg)} &
\multicolumn{1}{c}{} &
\multicolumn{1}{c}{(mag)} &
\multicolumn{1}{c}{} &
\multicolumn{1}{c}{} &
\multicolumn{1}{c}{(mag)} &
\multicolumn{1}{c}{} &
\multicolumn{1}{c}{} &
\multicolumn{1}{c}{(mag)} &
\multicolumn{1}{c}{} &
\multicolumn{1}{c}{} &
\multicolumn{1}{c}{(mag)} &
\multicolumn{1}{c}{} &
\multicolumn{1}{c}{} &
\multicolumn{1}{c}{(mag)} &
\multicolumn{1}{c}{} 
\\ 
\hline  
\hline
\dots        & \dots       && \dots  & \dots  & \dots  & \dots  & \dots  & \dots  & \dots  & \dots  & \dots  & \dots  & \dots  & \dots  & \dots  & \dots  \\
 $170.09283$  &  $12.88126$   &&   $14.40$  &  $0.04$  &  $0.09$  &  $ 13.75$  &  $0.10$  &  $0.09$  &  $ 13.03$  &  $0.06$  &  $0.03$  &  $ 12.14$  &  $0.04$  &  $-0.03$  &  $  8.28$  &  $0.01   $  \\  
 $170.05499$  &  $13.00885$   &&   $13.79$  &  $0.16$  &  $-0.09$  &  $ 13.76$  &  $0.13$  &  $0.15$  &  $ 11.11$  &  $0.11$  &  $0.02$  &  $  9.52$  &  $0.10$  &  $0.21$  &  $  4.51$  &  $0.05   $  \\  
 $170.06840$  &  $12.96378$   &&   $14.03$  &  $0.13$  &  $1.36$  &  $ 13.76$  &  $0.12$  &  $1.54$  &  $ 11.51$  &  $0.13$  &  $1.71$  &  $  9.79$  &  $0.08$  &  $1.85$  &  $  3.68$  &  $0.01   $  \\  
 $170.01328$  &  $12.92316$   &&   $13.89$  &  $0.05$  &  $-0.02$  &  $ 13.80$  &  $0.09$  &  $-0.05$  &  $ 13.86$  &  $0.05$  &  $0.10$  &  $ 13.61$  &  $0.06$  &  $-0.21$  &  $ 13.06$  &  $99.99   $  \\  
 $170.05455$  &  $13.00498$   &&   $14.12$  &  $0.10$  &  $0.15$  &  $ 13.81$  &  $0.12$  &  $0.24$  &  $ 11.63$  &  $0.08$  &  $99.99$  &  $  9.71$  &  $0.08$  &  $99.99$  &  $  4.78$  &  $0.09 $  \\  
\dots        & \dots       && \dots  & \dots  & \dots  & \dots  & \dots  & \dots  & \dots  & \dots  & \dots  & \dots  & \dots  & \dots  & \dots  & \dots  \\
\hline  
\hline  
\end{tabular} 
\end{small} 
\end{center}  
\end{table}

\begin{table}   
\begin{center}   
\begin{small}  
\caption{Catalog for $3,548$ Point Sources in NGC\,$3184$} 
\label{tab:n3184}   
\begin{tabular}{rrrrrrrrrrrrrrrrr}   
\\   
\hline  
\hline  
\multicolumn{1}{c}{RA} &
\multicolumn{1}{c}{Dec} &
\multicolumn{1}{c}{} &
\multicolumn{1}{c}{$m_{3.6}$} &
\multicolumn{1}{c}{$\sigma_{3.6}$} &
\multicolumn{1}{c}{$\delta_{3.6}$} &
\multicolumn{1}{c}{$m_{4.5}$} &
\multicolumn{1}{c}{$\sigma_{4.5}$} &
\multicolumn{1}{c}{$\delta_{4.5}$} &
\multicolumn{1}{c}{$m_{5.8}$} &
\multicolumn{1}{c}{$\sigma_{5.8}$} &
\multicolumn{1}{c}{$\delta_{5.8}$} &
\multicolumn{1}{c}{$m_{8.0}$} &
\multicolumn{1}{c}{$\sigma_{8.0}$} &
\multicolumn{1}{c}{$\delta_{8.0}$} &
\multicolumn{1}{c}{$m_{24}$} &
\multicolumn{1}{c}{$\sigma_{24}$} 
\\
\multicolumn{1}{c}{(deg)} &
\multicolumn{1}{c}{(deg)} &
\multicolumn{1}{c}{} &
\multicolumn{1}{c}{(mag)} &
\multicolumn{1}{c}{} &
\multicolumn{1}{c}{} &
\multicolumn{1}{c}{(mag)} &
\multicolumn{1}{c}{} &
\multicolumn{1}{c}{} &
\multicolumn{1}{c}{(mag)} &
\multicolumn{1}{c}{} &
\multicolumn{1}{c}{} &
\multicolumn{1}{c}{(mag)} &
\multicolumn{1}{c}{} &
\multicolumn{1}{c}{} &
\multicolumn{1}{c}{(mag)} &
\multicolumn{1}{c}{} 
\\ 
\hline  
\hline
\dots        & \dots       && \dots  & \dots  & \dots  & \dots  & \dots  & \dots  & \dots  & \dots  & \dots  & \dots  & \dots  & \dots  & \dots  & \dots  \\
 $154.53453$  &  $41.39239$   &&   $16.82$  &  $0.10$  &  $0.22$  &  $ 16.13$  &  $0.06$  &  $0.13$  &  $ 15.70$  &  $0.06$  &  $0.04$  &  $ 14.37$  &  $0.12$  &  $-0.14$  &  $ 10.91$  &  $0.14   $  \\  
 $154.63606$  &  $41.39248$   &&   $16.35$  &  $0.09$  &  $0.19$  &  $ 16.13$  &  $0.04$  &  $0.10$  &  $ 16.29$  &  $0.07$  &  $0.46$  &  $ 15.34$  &  $0.07$  &  $-0.50$  &  $ 12.46$  &  $99.99   $  \\  
 $154.67640$  &  $41.38895$   &&   $16.47$  &  $0.12$  &  $0.29$  &  $ 16.14$  &  $0.09$  &  $0.21$  &  $ 15.76$  &  $0.09$  &  $0.33$  &  $ 14.63$  &  $0.05$  &  $-0.22$  &  $  99.99$  &  $99.99   $  \\  
 $154.56665$  &  $41.44029$   &&   $16.09$  &  $0.12$  &  $0.39$  &  $ 16.15$  &  $0.12$  &  $0.71$  &  $ 13.72$  &  $0.11$  &  $0.87$  &  $ 11.16$  &  $0.03$  &  $0.32$  &  $  7.93$  &  $0.06   $  \\  
 $154.39740$  &  $41.38107$   &&   $16.75$  &  $0.15$  &  $1.08$  &  $ 16.15$  &  $0.08$  &  $0.79$  &  $ 15.20$  &  $0.19$  &  $99.99$  &  $ 14.70$  &  $0.27$  &  $0.02$  &  $  99.99$  &  $99.99 $  \\
\dots        & \dots       && \dots  & \dots  & \dots  & \dots  & \dots  & \dots  & \dots  & \dots  & \dots  & \dots  & \dots  & \dots  & \dots  & \dots  \\
\hline  
\hline  
\end{tabular} 
\end{small} 
\end{center}  
\end{table}

\end{landscape}

\clearpage

\end{appendix}

\end{document}